\documentclass[superscriptaddress,amsmath,amssymb,prd,preprintnumbers,showpacs,twocolumn,nofootinbib]{revtex4-2}
\usepackage[colorlinks=true,pdfstartview=FitV,linkcolor=blue,citecolor=blue,urlcolor=blue,breaklinks=true]{hyperref}
\usepackage{graphicx}
\usepackage[T1]{fontenc} 
\usepackage{float} 
\usepackage{amssymb,amsmath,bm,natbib}
\usepackage{color}
\usepackage{slashed}
\usepackage{graphics}
\usepackage{graphicx}
\usepackage[utf8]{inputenc}
\usepackage[caption=false]{subfig}
\usepackage{hyperref}
\usepackage{url}
\usepackage{dsfont}
\usepackage{float}
\usepackage{cancel}
\usepackage{units}
\usepackage{blindtext}
\usepackage[utf8]{inputenc}
\usepackage{upgreek}
\usepackage{booktabs}
\usepackage[dvipsnames,table,xcdraw]{xcolor}
\usepackage{enumerate}
\usepackage{mathtools}
\usepackage{soul,color}
\usepackage[normalem]{ulem}
\newcommand{\orcid}[1]{\href{https://orcid.org/#1}{\includegraphics[width=10pt]{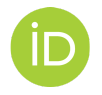}}}
\usepackage[T1]{fontenc} 
\usepackage{xcolor}
\usepackage{graphicx}
\usepackage{cleveref}
\usepackage{amsopn}
\usepackage{braket}
\usepackage{bbold}

\def \be {\begin{equation}}
\def \ee {\end{equation}}
\def \bea {\begin{eqnarray}}
\def \eea {\end{eqnarray}}
\def \sla {\slashed}
\def \be {\begin{equation}}
\def \ee {\end{equation}}
\def \bea {\begin{eqnarray}}
\def \eea {\end{eqnarray}}
\def \sla {\slashed}
\def \slsh {\slashed}
\begin{document}
\title{Higher-order Chern-Simons extensions to QED in $2+1$ dimensions  }
\author{Ricardo Avila\orcid{0000-0002-3488-6503}}
\email{  raavilavi@gmail.com    }
\affiliation{Escuela de Fonoaudiología, Facultad de Ciencias de la 
Salud, Universidad Católica Silva Henríquez, 
Chile.}
 \author{Albert Yu. Petrov\orcid{0000-0003-4516-655X}}
\email  { petrov@fisica.ufpb.br }
\affiliation{Departamento de F\'{\i}sica, Universidade Federal da Para\'{\i}ba,
 Caixa Postal 5008, 58051-970, Jo\~ao Pessoa, Para\'{\i}ba, Brazil. }
 \author{Carlos M. Reyes\orcid{0000-0001-5140-6658}}
\email{creyes@ubiobio.cl}
\affiliation{Centro de Ciencias Exactas, Facultad de Ciencias, 
Universidad del B\'{i}o-B\'{i}o, Chill\'{a}n, Chile.}
\author{C\'esar Riquelme\orcid{0000-0003-0837-3891}}
\email{ceriquelme@udec.cl}
\affiliation{Departamento de F\'{i}sica, Universidad de 
Concepci\'{o}n, Concepci\'on, Casilla 160-C, Chile.}
\affiliation{Centro de Ciencias Exactas, Facultad de Ciencias, 
Universidad del B\'{i}o-B\'{i}o, Chill\'{a}n, Chile.}
  \author{Angel Sanchez\orcid{0000-0002-8237-5257}}
\email{ ansac@ciencias.unam.mx }
\affiliation{Departamento de F\'isica, Facultad de Ciencias, 
Universidad Nacional Aut\'onoma de M\'exico,
Apartado Postal 70-542, Ciudad de M\'exico 04510, Mexico.}
\begin{abstract}
In this work, we investigate radiative corrections in the 
higher-order extension of the
Maxwell-Chern-Simons model coupled to standard spinor matter in $2+1$ dimensions.
We begin analyzing the higher-order gauge sector, where we find 
the modes and the polarizations vectors associated to a massive photon and ghost field.
The higher-order gauge model is canonically quantized and as expected 
the resulting algebra of creation and annihilation operators
corresponds to an indefinite metric in Hilbert space.
Subsequently, we compute all relevant one-loop diagrams in the modified QED
starting with the fermion self-energy. We show that the induced corrections 
to the fermion two-point function 
produce two independent fermionic degrees of freedom, which 
can be included in a redefined Lagrangian describing two decoupled fermions fields, 
one corresponding to a physical particle and the other to a negative-norm ghost state.
We take advantage of this decomposition to compute the photon polarization operator and the
vertex correction, both of which are found to be finite. Finally, we analyze
the causal behavior of the model by computing the commutator 
  of gauge fields at different spacetime points, and found that microcausality is preserved.
\end{abstract}
\maketitle
\section{Introduction}
Chern-Simons (CS) theory is a topological quantum field theory 
that arises naturally in odd-dimensional spacetimes. 
It was first derived as an effective contribution from the high-temperature 
limit of QCD and electroweak models  
in four dimensions~\cite{Linde:1978px,Gross:1980br,Efraty:1992gk}.
The CS theory provides a rich framework for exploring a wide range of physical phenomena.
In particular, the CS term generates a gauge-invariant mass for the gauge field  
in $2+1$ spacetime dimensions~\cite{Schonfeld:1980kb,Deser:1981wh}.
Moreover, 
 the definitions of various
topological invariants, such as the Jones polynomial in knot theory,
can be connected with the non-abelian CS action~\cite{Witten:1988hf}, revealing
a deep interplay between topological aspects of quantum field theory and lower dimensional conformal field 
theory. 
Furthermore, the CS
term presents interesting possibilities in describing various physical
systems, as the quantum Hall effect~\cite{Girvin:1987fp,Zhang:1988wy,Hansson:2016zlh,Iengo:1991zbc}, 
topological materials including topological 
insulators~\cite{Hasan:2010xy,vonDossow:2025bwr}, 
spectroscopic features in Weyl simimetals~\cite{TaAs,TaAs2,Martin-Ruiz:2018ets,Kostelecky:2021bsb}, also
extensions of gravity~\cite{Deser:1981wh,Achucarro:1986uwr},
vortex dynamics~\cite{Vortices,Casana:2014dfa}, 
Skyrme models~\cite{Casana:2019zlb}, Higgs mechanism~\cite{Bazeia:2012ux},
topological gravity~\cite{Merino}, 
supersymmetric CS~\cite{Lehum:2025vkw} and anisotropic QED~\cite{CS_Lifshitz};
see also the reviews~\cite{Dunne:1998qy,Alexander:2009tp}.

A topic that has attracted growing interest 
 over the years is the 
quantum behavior of the abelian CS term. 
Quantum corrections 
to modified QED incorporating the CS term exhibit several noteworthy features, 
 such as 
the smoothness of infrared divergencies 
resulting from the massive gauge mode generated by the gauge-invariant CS term.
Moreover, in modified QED the polarization tensor receives only 
one-loop corrections, with all higher-order contributions 
vanishing due to a remarkable cancellation~\cite{Chen}, a result known as 
the Coleman--Hill theorem~\cite{C-H}. 
The CS term has also been studied to extend to parity violating mass 
terms~\cite{Bashir:2008ej,Concha-Sanchez:2013wre}.

The higher-derivative extension of the CS term was proposed some 
years ago by S.~Deser and R.~Jackiw in order to describe the 
momentum expansion of the effective action in three-dimensional QED~\cite{JTD}. 
Some works on such an extension have focused on relic 
symmetries~\cite{Testing}, thermal effects~\cite{Ayala:2010fm,petrov}, Lorentz-violating planar
electrodynamics~\cite{Ferreira:2004hx,Lisboa-Santos:2023pwc,Borges:2024vjc,Rocha:2024ndf}, analysis of 
unitarity~\cite{Avila:2019xdn,Ferreira:2020wde}, and vector multiplets~\cite{Kaparulin:2021kle}.
It is well known that higher time derivatives can improve the ultraviolet 
behavior of a theory by introducing higher powers of momentum 
in the denominators of propagators. However, they may also lead to an indefinite metric, with additional degrees of freedom associated to ghost fields~\cite{LW1,LW2}. 
 In four dimensions, one can consider an analogue, which, however, is not topological in essence 
 and necessarily breaks Lorentz invariance~\cite{Jackiw:2003pm}.
In this work, we explore the higher-order CS term in $2+1$
dimensions, with particular emphasis on quantum corrections induced by ghost states associated with the indefinite metric of the theory.

The work is organized as follows. In Sec.~\ref{sectionII}, we present the QED model,  
finding the  gauge propagator together with its polarization vectors.
In Sec.~\ref{sectionIII}, we canonically quantize the model
and derive the commutator algebra of creation and annihilation operators. 
In Sec.~\ref{sectionIV} we compute 
the radiative correction up to one-loop level of the fermion 
self-energy, photon polarization operator and vertex function.
Also, we analyze the superficially 
degree of divergence of each one-loop diagram.
In Sec.~\ref{sectionV},  we study microcausality and prove that 
it is preserved. Finally, in Sec.~\ref{sectionVI} we give our final remarks. In 
Appendix~\ref{App:A} we present the calculation of an element of the commutator of gauge fields
in a general frame.
\section{Extended QED model in $(2+1)$-dimensions   }\label{sectionII}
Our model is based on the higher-derivative QED Lagrangian
in $2+1$ dimensions, given by
\begin{align}\label{Lagrangian_parts}
\mathcal{L}&=\bar \psi \left(i\sla{\partial}+ e \sla{A}-
m \right)\psi-\frac{1}{4\gamma}F_{\mu\nu}
F^{\mu\nu}  -\frac{1}{2\xi}(\partial_\mu A^{\mu})^2\nonumber 
\\  &\hspace{4em}+ \mathcal L_{\mathrm {CSe}}  \,,
\end{align}
where $\mathcal{L}_{\mathrm{CSe}}$ includes both the CS 
term and its higher-derivative extension~\cite{JTD}, written as
\begin{align}\label{Lagrangian_2}
	\mathcal{L}_{\mathrm{CSe}}&= \frac{1}{2}\epsilon^{\alpha\beta\gamma} 
    A_\alpha  \left(  \mu+g \Box  \right)
	\partial_\beta  A_\gamma \,.
\end{align}
In the Lagrangian~\eqref{Lagrangian_parts}, we have 
coupled the photons with 
standard fermions and considered 
the usual interaction term.
Also, we have included 
 a covariant gauge-fixing term with $\xi>0$ and
 a constant $\gamma$ in the standard Maxwell term,
 which allow us 
 to obtain the pure CS theory by taking the limit $\gamma \to \infty$ together 
 with $g\to 0$. Let us emphasize that the CS term has been included from the outset, as it is 
naturally generated from the standard photon polarization operator diagram~\cite{Chen}. 

Taking into account that the action is dimensionless, we find that 
the mass dimensions of the parameters and fields are
\begin{align}
\left[ \mu \right]&=1 \,, \quad [g]=-1 \,, \quad [e]=\frac{1}{2} \,, 
\nonumber \\  [\psi]&=1  \,,  \quad  [A_{\mu}]=\frac{1}{2}\, ,
\end{align}
while the constants $\gamma$ and $\xi$
are dimensionless.
 
In this work we employ the mostly minus sign convention for the metric signature, i.e.,
 $ \eta^{\mu \nu}=\text{diag}(1,-1,-1)$,  together with
  the Levi-Civita convention 
$\epsilon^{012}=1$ and 
the two-dimensional realization of the Dirac algebra
\begin{align}
 \gamma^0&=\sigma^3  =\left(\begin{array}{cc}1 & 0 \\ 0 & -1\end{array}\right)  
   \,,  \nonumber \\    \gamma^1&=i\sigma^1 =\left(\begin{array}{cc}0 & i \\i & 0\end{array}\right) \,, \nonumber 
  \\  \gamma^2&=i\sigma^2 =\left(\begin{array}{cc}0 & 1 \\-1 & 0\end{array}\right)    \,,
\end{align}
with $\sigma^1, \sigma^2,\sigma^3 $ being the Pauli matrices.
Furthermore, one can check that the Dirac matrices satisfy the Clifford algebra
\begin{align}\label{diracalgebra}
\{  \gamma^{\mu}, \gamma^{\nu}  \}&=2\eta ^{\mu \nu}  \,,
\end{align}
 and
 the relations
\begin{align}\label{relationgamma}
 \gamma^{\mu}  \gamma^{\nu}   &= \eta ^{\mu \nu }  
 \mathbb{1}_{2\times 2}-i\epsilon^{\mu \nu \alpha}  \gamma_{\alpha} \,,  \\ 
  \mathrm{tr} (\gamma^{\mu} \gamma^{\nu}\gamma^{\rho}  )&=-2i \epsilon^{\mu \nu \rho}\,,
\notag \\ \notag
\gamma^{\mu} & = \eta^{\mu \nu} \gamma_{\nu}  \,,
\end{align}
 where $\mathbb{1}_{2\times 2}$ is the $2\times 2$ unit matrix.
\subsection{The gauge propagator  }\label{subsecII-B}
Let us consider the free Lagrangian in the gauge sector:
\begin{align}\label{lagrangianA}
&\mathcal L_A= -\frac{1}{4\gamma}F_{\mu\nu}
F^{\mu\nu}  -\frac{1}{2\xi}(\partial_\mu A^{\mu})^2\nonumber 
\\  &\hspace{3em}+
	\frac{1}{2}\epsilon^{\alpha\beta\gamma} 
    A_\alpha  \left(  \mu+g \Box  \right)
	\partial_\beta  A_\gamma \,.
\end{align}

The generalized Euler-Lagrange equation for the second-order 
Lagrangian $\mathcal L_A$ can be written as
\begin{align}\label{gen_EL}
-\partial_{\kappa} \partial_{\lambda}  \frac{\partial \mathcal L_A}
{\partial( \partial_{\kappa} \partial_{\lambda}  A_{\sigma})  }
+ \partial_{\rho}  \frac{\partial \mathcal L_A}
{\partial( \partial_{\rho}   A_{\sigma})  }-   \frac{\partial \mathcal L_A}
{\partial  A_{\sigma}  }=0\,.
\end{align}
From the Lagrangian~\eqref{lagrangianA} and 
using~\eqref{gen_EL}, we have the equation of motion 
\begin{align}\label{G-F-E}
&\Big[\frac{\eta^{\mu \nu}}{\gamma}\Box  
-\left(\frac{1}{\gamma}-\frac{1}{\xi}\right)\partial^\mu \partial^\nu 
\nonumber \\
&\hspace{5em} +\epsilon^{\mu \beta\nu} \left( \mu+ g\Box \right)
\partial_\beta    \Big]  A_\nu (x) =0\,.
\end{align}
Contracting Eq.~(\ref{G-F-E}) with $\partial_{\mu}$ yields
\begin{equation}\label{guage_eq}
\frac{1}{\xi}\Box \left(\partial \cdot A \right)=0\,,
\end{equation}
which by imposing suitable boundary conditions 
at infinity gives the condition
$\partial \cdot A=0$.

Now, we go to momentum space expanding
the gauge field in Fourier modes
\begin{align}
A_{\mu}(x)=\int \mathrm{d}^3 k A_{\mu} (k) \, e^{-ik\cdot x}\,.
\end{align} 
Replacing in the equation of motion~\eqref{G-F-E}, we obtain
in momentum space 
\begin{align}
S^{\mu \nu} (k)A_{\nu}(k)  =0\,,
\end{align}  
where we have defined the operator 
\begin{align}
S^{\mu \nu} (k)&= -\frac{k^2}{\gamma} 
\eta^{ \mu \nu } +\left(\frac 1 \gamma - \frac{1}{\xi}\right)  k^{\mu}  
	k^{\nu} 
  \notag \\ &\hspace{8em}   -i  \mathcal M(k)   \epsilon^{\mu \beta \nu}   k_{\beta}    \,,
\end{align}  
and we have introduced the function of momentum 
\begin{align}
\mathcal M(k) := \mu-g k^2\,,
\end{align}
which has dimension of mass.

By inverting the operator we find the 
propagator 
\begin{align}   \label{Propa}  
G_{\nu\rho}(k)=\frac{-i\gamma}{k^2- \gamma^2\mathcal{M}^2(k)}T_{\nu\rho}(k) \,,
\end{align}
which satisfies $S^{\mu \nu} 
G_{\nu \rho}=  i \delta ^{\mu}_{\rho}$,
with the definition
\begin{align} \label{Tmunu}
 T_{\nu\rho}(k)&:= \eta_{\nu\rho}-\left[1-\frac{\xi}{\gamma}
 \left(1- \frac{\gamma^2\mathcal{M}^2}{k^2}\right)\right]  \frac{k_\nu k_\rho}{k^2}  \nonumber \\
 &\hspace{6em}  
-i\gamma\mathcal{M}\epsilon_{\nu\beta\rho}\frac{k^\beta}{k^2} \,.
\end{align}
The propagator in the Landau gauge setting $\xi=0$, reads  
\begin{align}\label{Landau-Propa}
\Delta_{\nu \rho}(k)=  \frac{-i \gamma}{k^2-\gamma^2 \mathcal M^2 
}  \left( \eta_{ \nu \rho }- \frac{k_{\nu} 
  k_{\rho} }{k^2}    -i \gamma  \mathcal M 
  \epsilon_{\nu \beta \rho} \frac{k^{\beta}}{k^2} \right)\,,
\end{align}
which avoids infrared divergencies and will prove most convenient to use
in the subsequent sections.

The dispersion equation can be read off from 
the pole structure of the propagators in 
Eqs.~\eqref{Propa} and~\eqref{Landau-Propa}, which we write as
\begin{align}\label{Disp-Rel}
k^2-\gamma ^2\mathcal M^2=0\,.
\end{align}
Solving~\eqref{Disp-Rel} yields two massive propagating modes
\begin{subequations}
\begin{align} \label{omega}
 \omega_1= &\sqrt{\vec k^2+m_1^2}  \,, 
\\
W_2=& \sqrt{\vec k^2+M_2^2}  \label{W} \,,
\end{align}
 \end{subequations}
with masses
\begin{subequations}
\begin{align} \label{bosonmasses1}
m_1= &\frac{\sqrt{1+4\gamma^2 \mu g}-1}{2 \gamma g} \,,  \\ \label{bosonmasses2}
M_2=& \frac{\sqrt{1+4 \gamma^2 \mu g}+1}{2\gamma g} \,.
\end{align}
 \end{subequations}
 Both 
 solutions are guaranteed to be real, ensuring the 
 absence of tachyons in the theory. Also, the first solution 
$\omega_1$ corresponds to a physical massive photon, while the second solution $W_2$
 corresponds to a massive ghost state. This explicitly demonstrates the indefinite 
 metric structure of the theory~\cite{LW1,LW2} as one can show by taking 
 the low energy limit $g\to 0$. Hence, expanding our dispersion
relations up to the second order in $g$, we have 
\begin{subequations}
\begin{align}
\omega_1 &\approx  \sqrt{ \vec k^2+ \gamma^2 \mu^2}- 
\frac{g \gamma^4\mu^3}{ \sqrt{ \vec k^2+\gamma^2 \mu^2} }+\mathcal O(g^2)\,,
\\
 W_2 &\approx \frac{1}{\gamma g}+ \gamma \mu +g \left( 
 \frac{\vec k^2 \gamma}{2}-\gamma^3\mu^2 \right)+\mathcal O(g^2)\,,
\\ \label{part_mass}
m_1 &\approx  \mu \gamma    -   \mu^2  \gamma^3  g+  \mathcal O(g^2)\,,
\\ \label{part_Mass}
 M_2 &\approx  \frac{1 }{ \gamma g} +\mu \gamma -   \mu^2  \gamma^3  g+  \mathcal O(g^2)  \,,
\end{align}
\end{subequations}
and we can see that in the limit $g\to 0$, the physical solution 
remains regular, as expected, while 
the ghost solution exhibits a singularity, consistent 
with its problematic nature; in particular connected to stability and causality.

Furthermore, the equations~\eqref{part_mass} and~\eqref{part_Mass} 
reveal a significant mass gap between ghost and particle, 
given by $M_2 - m_1 = 1/(\gamma g)$. This gap disappears in the pure Chern-Simons limit 
by taking the limit $\gamma \to \infty$, at which both
masses coincide $M_2 = m_1$. 
Throughout this work, 
 and without loss of generality, we assume both $g > 0$ and $\gamma > 0$.
\subsection{Polarization vectors}\label{subsecII-C}
We begin by defining an orthogonal basis of $(2+1)$-dimensional
Minkowski spacetime, given by the real vectors
\begin{subequations}
  \begin{align}
    e^{(0)\mu}&=\frac{1}{\sqrt{k^2}}k^\mu \,,   \\
    e^{(1)\mu}&=\frac{1}{\sqrt{G}}\epsilon^{\mu\beta\gamma}k_\beta n_\gamma \,, \\
    e^{(2)\mu}&=-\frac{1}{\sqrt{k^2}}\epsilon^{\mu\beta\gamma}k_\beta e_\gamma^{(1)}
 \notag \\& =\frac{1}{\sqrt{k^2G}}\left(k^2 n^\mu-k^\mu (k\cdot n)\right) \,, 
\end{align}  
\end{subequations}
where $G\equiv(k\cdot n)^2-k^2n^2$ and $n^\mu$ is an auxiliary field.

The three vector
basis $ e^{(a)\mu}$ are normalized
according to
\begin{align}
    e^{(a)}\cdot e^{(b)}=g_{ab}\,,
\end{align}
with $a=0,1,2$ and $g_{ab}=\text{diag.}(1,-1,-1)$.
They also satisfy
\begin{align}
    \epsilon^{\mu\beta\gamma}k_\beta e_\gamma^{(2)}&=\sqrt{k^2}e^{(1)\mu}\,, \\
    \epsilon^{\mu\beta\gamma}k_\beta e_\gamma^{(1)}&=-\sqrt{k^2}e^{(2)\mu}\,.
\end{align}
We define the complex basis
\begin{subequations}
   \begin{align}
    \varepsilon^{(0)\mu}&=e^{(0)\mu}\,, \\
    \varepsilon^{(+)\mu}&=\frac{1}{\sqrt{2}}\big(e^{(2)\mu}+ie^{(1)\mu}\big)\,, \\
    \varepsilon^{(-)\mu}&=\frac{1}{\sqrt{2}}\big(e^{(2)\mu}-ie^{(1)\mu}\big)\,,
\end{align}  
and introduce the index $\lambda=0,\pm$.
\end{subequations}
Here the $\pm$ modes are orthogonal to the momentum, i.e.,
$k\cdot \varepsilon^{(\pm)}=0$. The vectors defining the
complex basis satisfy the relations
\begin{eqnarray}
    \varepsilon^{(\lambda)}\cdot \varepsilon^{(\lambda')*}&=&g_{\lambda \lambda'}\,, \\ 
    \epsilon^{\mu\beta\sigma}k_\beta \varepsilon_\sigma^{(\pm)}&=&\mp i\sqrt{k^2}\varepsilon^{(\pm)\mu} \,,
\end{eqnarray}
with $g_{\lambda \lambda'}=\text{diag.}(1,-1,-1)$.

This set of complex basis vectors diagonalizes the $S^{\mu\nu}$ as follows
\begin{subequations}
 \begin{eqnarray}
    S^\mu_{\phantom{\mu}\nu}(p)\varepsilon^{(0)\nu}&=&\Lambda_0(p)\varepsilon^{(0)\mu}\,,\\
    S^\mu_{\phantom{\mu}\nu}(p)\varepsilon^{(+)\nu}&=&\Lambda_+(p)\varepsilon^{(+)\mu}\,, \\
    S^\mu_{\phantom{\mu}\nu}(p)\varepsilon^{(-)\nu}&=&\Lambda_-(p)\varepsilon^{(-)\mu}\,,
\end{eqnarray}   
\end{subequations}
where
\begin{subequations}
 \begin{align}
    \Lambda_0(k)&=-\frac{k^2}{\xi}\,, \\
    \Lambda_+(k)&=-\frac{k^2}{\gamma}-\sqrt{k^2} \mathcal{M}(k)\,, \label{lambda+}\\
    \Lambda_-(k)&=-\frac{k^2}{\gamma}+\sqrt{k^2}\mathcal{M}(k)\,. \label{lambda-}
\end{align}   
\end{subequations}
We notice that $W_2$ is a solution related to $\Lambda_+(W_2,\vec k)=0$ 
while $\omega_1$ is related to $\Lambda_-(\omega_1,\vec k)=0$. 

The dispersion equation follows from the product of the three eigenvalues of $S^{\mu\nu}$
\begin{align}
    \prod_{\lambda=0,\pm}\Lambda_\lambda(p)
    =-\frac{(k^2)^2}{\xi \gamma^2}\bigg(k^2- \gamma^2\mathcal{M}^2(k)\bigg)=0\,.
\end{align}
One can show that
\begin{eqnarray}
    \varepsilon_\mu^{(\pm)}\varepsilon_\nu^{(\pm)*}   &=&-\frac{1}{2}\left(\eta_{\mu\nu}-\frac{k_\mu k_\nu}{k^2}\pm 
    i\epsilon_{\mu\beta\nu}\frac{k^\beta}{\sqrt{k^2}}\right)\,,
\label{relationspol}
\end{eqnarray}
from where we make the connection with the propagator, by calculating
\begin{align}    &\sum_{\lambda,\lambda'=0,\pm}
g_{\lambda\lambda'}\frac{\varepsilon_\mu^{(\lambda)}\varepsilon_\nu^{(\lambda')*}}{\Lambda_\lambda} 
=-iG_{\mu\nu}.
\end{align}
We can also make a connection with the equation of motion operator in  momentum space 
\begin{align}
&\sum_{\lambda,\lambda'=0,\pm}g_{\lambda\lambda'}\Lambda_\lambda
\varepsilon_\mu^{(\lambda)}\varepsilon_\nu^{(\lambda')*}  =S_{\mu\nu}(\xi,k)\,,
\end{align}
Thus the set $\lbrace \varepsilon^{(\lambda)\mu}\rbrace$ 
correspond to independent eigenvector
solutions of the equation of motion operator. 
\section{Canonical quantization }\label{sectionIII}
According to Ostrogradsky variational formalism~\cite{Ostro},
the momentum variables in our model can be written as
\begin{align}
    P^\mu&:=\frac{\partial \mathcal{L}_A}{\partial \dot{A}_\mu}-\frac{\partial \Pi^\mu}{\partial t}\,,
    \\
    \Pi^\mu&:=\frac{\partial \mathcal{L}_A}{\partial \ddot{A}_\mu}\,.
\end{align}
The explicit form of momentum variables looks like
\begin{align}
    P^\mu&=-\frac{1}{\gamma}
F^{0\mu}   -\frac{1}{\xi}(\partial_\lambda A^{\lambda})\eta^{0\mu} +
\frac{1}{2}\epsilon^{\alpha 0 \mu}    \big(\mu+g\Box \big)A_\alpha \nonumber
\\ & \hspace{4em} - \frac{g}{2}\epsilon^{\mu\beta\gamma}       \partial_\beta \dot{A}_\gamma \,,
\\
 \Pi^\mu  &= \frac{g}{2}\epsilon^{\mu\beta\gamma} \partial_\beta  A_\gamma\,.
\end{align}
We impose the equal-time commutation relations on the canonical variables
\begin{align} \label{ETCR}
\big[A_\mu(t,\vec{x}),P_\nu(t,\vec{y})\big]&=i\eta_{\mu\nu}\delta^{(2)}(\vec{x}-\vec{y})\,, \\ \label{ETCR2}
\big[\dot{A}_\mu(t,\vec{x}),\Pi_\nu(t,\vec{y})\big]&=i\eta_{\mu\nu}\delta^{(2)}(\vec{x}-\vec{y})\,,
\end{align}
where all the others are defined to vanish.
As shown in Refs.~\cite{Testing,Avila:2019xdn},
it is very likely that constraints will
arise,
which in turn lead to the appearance of Dirac 
brackets~\cite{Hanson:1976cn,Mukherjee:2011da,Sararu:2014sua}. A detailed analysis of this
issue is left for future work, as it lies beyond the scope of the present study.

Let us split the gauge field in terms of the massive degrees
of freedom corresponding to a massive photon and a massive ghost field respectively, as follows
\begin{align}\label{gauge_decomposition}
    A_\mu(x)=\bar{A}_\mu(x)+G_\mu(x)\,.
\end{align}
We see that the wave-vector in the expansion must be a solution to the 
following eigenvector equation:
\begin{align}
    S^{\mu}_{\phantom{\mu}\nu}(k)\varepsilon^{(\lambda)\nu}(k)
    =\Lambda_\lambda(k) \varepsilon^{(\lambda)\nu}(k)=0\,.
\end{align}
Since $\bar{A}_\mu(x)$ is a massive photon with the positive
energy $\omega_1$ and mass $m_1$ satisfying 
$\Lambda_-(\omega_1)=0$ we conclude that it belongs 
to the subspace $(-)$. Proceeding similarly for the 
ghost with $\Lambda_+(W_2)=0$,
we conclude that it is related to the $(+)$ subspace. 
Furthermore, we notice that $\bar{A}_\mu$ and $G_\mu$ satisfy 
the orthogonality condition for the $A_\mu$ field, $k^\mu \bar{A}_\mu= k^\mu G_\mu=0$.
Taking all this into account,
we arrive at an expansion of both physical 
photon and ghost orthogonal to the direction of propagation $k^\mu$ 
\begin{widetext}
\begin{eqnarray}
\bar{A}_\mu(x)&=&\int \frac{\mathrm{d}^2
    \vec{k}}{(2\pi)^2}\frac{1}
    {\Lambda_-'(\omega_1,\vec {k})}\bigg[\varepsilon_\mu^{(-)}(k_0,\vec{k})
    a^{(-)}_{\vec{k}}e^{-i k\cdot x}+ \varepsilon_\mu^{(-)*}
    (k_0,\vec{k})a^{(-)\dagger}_{\vec{k}}e^{i k\cdot x}   \bigg]_{k_0=\omega_1} \,, \\
    G_\mu(x)&=&\int \frac{\mathrm{d}^2\vec{k}}{(2\pi)^2}
    \frac{1}{\Lambda_+'(W_2,\vec{k})}\bigg[\varepsilon_\mu^{(+)}
    (k_0,\vec{k})b^{(+)}_{\vec{k}}e^{-i k\cdot x}+ \varepsilon_\mu^{(+)*}(k_0,\vec{k})
    b^{(+)\dagger}_{\vec{k}}e^{i k\cdot x}   \bigg]_{k_0=W_2} \,,
\end{eqnarray}
\end{widetext}
where we used $\Lambda_\lambda'(k)=\partial_{k_0}\Lambda_\lambda(k)$ 
as normalization in order to define a Lorentz 
invariant measure~\cite{Colladay:2016rsf}.

From the ETCR~\eqref{ETCR} \eqref{ETCR2},
we impose the creation and annihilation operators algebra
\begin{eqnarray}
    \big[a^{(-)}_{\vec{p}},a^{(-)\dagger}_{\vec{k}}\big]&=&-(2\pi)^2
    \Lambda_-'(\omega_1,\vec{k})\delta^{(2)}(\vec{p}-\vec{k})\,,\\
    \big[b^{(+)}_{\vec{p}},b^{(+)\dagger}_{\vec{k}}\big]&=&
    -(2\pi)^2\Lambda_+'(W_2,\vec{k})\delta^{(2)}(\vec{p}-\vec{k})\,,
    \label{algebrabop}
\end{eqnarray}
with all other commutators being zero and 
\begin{eqnarray}\label{lambdas}
   \left. \Lambda_-^\prime (k)  \right|_{k_0=\pm \omega_1}    &=&\mp 2\omega_1
    \bigg(\frac{\sqrt{1+4\gamma^2\mu g}}{2\gamma}\bigg)\,, \\ \label{lambdas2}
    \Lambda_+' \left( k \right)   \vert_{k_0=\pm W_2}&=&\pm 
    2 W_2\bigg(\frac{\sqrt{1+4\gamma^2\mu g}}{2\gamma}\bigg) \,.
\end{eqnarray}
Note that the commutator~\eqref{algebrabop} admits 
negative-norm states, leading to an indefinite metric in Hilbert space.
\section{Radiative corrections}\label{sectionIV}
In this section, we compute the relevant one-loop radiative corrections
of the higher-order QED model in $(2+1)$ dimensions.
In the first subsection, we analyze the divergences of the model. In the next subsection,
we compute the 
fermion self-energy and find that it induces
a higher-order term proportional to $g\Box$ coming from the higher-order gauge sector.
In consequence, we redefine the fermionic Lagrangian in terms of two decoupled fermion fields.
In the two last subsections, we compute the vacuum polarization operator and the 
vertex diagram, by employing the modified fermion propagator.
\subsection{Analysis of divergences}\label{subsecIV-A}
Let us perform an analysis of the superficially divergent integrals
in our modified QED. 
Using the UV asymptotic for the gauge propagator 
in the Landau gauge~\eqref{Landau-Propa},
one has the superficial degree of divergence $D$ 
\begin{align}
 D=3L-3P_A-P_{\psi} =3+2P_{\psi}-3V \,,
\end{align}
where $L$ is the number of loops, and $P_A$ and $P_{\psi}$ 
are the numbers of gauge and spinor propagators respectively.
Here we used the topological identity $L+V-P=1$, together 
with the 
fact that the total number of propagators are $P=P_A+P_{\psi}$.

Then, we can relate numbers of vertices, propagators and external 
legs ($E_A$ for external photons and $E_{\psi}$ for external fermions)
calculating the numbers of all fields associated to all vertices. Since we have basically a QED theory 
with the standard gauge-spinor vertex, we have
\begin{align}
V=2P_A+E_A   \to P_A=\frac{1}{2}(V-E_A)\,,  \\
2V=2P_{\psi}+E_{\psi}   \to P_{\psi}=\frac{1}{2}(2V-E_{\psi})\,,
\end{align}
for the gauge and fermion fields respectively, relating the number of vertices, propagators and external legs.
This allows us to write $D$ as
\begin{equation}
D=3-E_{\psi}-V.
\end{equation}
From the formula for $D$ 
we find that in divergent diagrams, one has $E_{\psi} \leq 2$. In particular, we have for 
the photon polarization tensor graph $V=2, E_{\psi}=0$, and so $D=1$. 
We will see that within the dimensional 
regularization scheme this contribution is finite, see section~\ref{sectionIV}.
For the fermion one-loop self-energy graph, one has $V=2,\quad  E_{\psi}=2, \quad D=-1$ which
 is finite. The fermion-photon triple vertex function is also 
 finite, corresponding to $D=-2$.  
 All one-loop Feynman diagrams with more numbers of legs and vertices are finite as well. Performing 
 a general analysis we have that the mass dimension in our interacting theory is 
  $\Delta=1/2$~\cite{Schwartz:2014sze,Peskin:1995ev}. 
  
We conclude
that our theory is super-renormalizable and we can have divergences up to two loops,
by straightforward consideration of Feynman diagrams. However, by 
gauge symmetry reasons the only potentially divergent contribution from the gauge sector, that is,  $A^{\alpha}A_{\alpha}$ 
term cannot arise even in two loops whether 
corrections with derivatives (CS and Maxwell term) are finite in two loops. So, the only possible divergences
in principle  are the two-loop corrections to the spinor kinetic term and spinor-vector vertex.
\subsection{Fermion self energy}\label{subsectionIV-B}
Let us write down the two-point function for the spinor field:
\begin{align}\label{fermionselfP}
i\Sigma(p)&=(-ie)^2\int\frac{ \mathrm{d}^3k}{(2\pi)^3}
\gamma^{\alpha}S_0(p-k)\gamma^{\beta}\Delta_{\alpha\beta}(k)  \,.
\end{align}
We work in the Landau gauge~\eqref{Landau-Propa} and set
$\gamma=1$ ensuring
a nontrivial impact of the Maxwell term.

The fermion propagator is the standard one
\begin{align} \label{fermion-propagator}
S_0(k)&=\frac{i}{\slsh{k}-m} \,.
\end{align}
This expression can be written as a following sum
of four terms: 
\begin{align} 
\Sigma(p)&=\Sigma^{(1)}(p)+\Sigma^{(2)}(p)+\Sigma^{(3)}(p) 
+\Sigma^{(4)}(p)  \,,
\end{align}
where the first one is
\begin{subequations}
\begin{align} \label{part1}
\Sigma^{(1)}(p)&=  ie^2  \int\frac{\mathrm{d}^3k}{(2\pi)^3}
\gamma^{\mu}    \left(\frac{\sla{p}-\sla{k}+m}{(p-k)^2-m^2}  \right) 
 \gamma_{\mu}\notag \\ & \hspace{8em} \times
 \frac{1}{k^2-\mathcal M ^2(k)}  \,,
\end{align} 
the second one
\begin{align}
\Sigma^{(2)}(p)&=  -ie^2  \int\frac{\mathrm{d}^3k}{(2\pi)^3}
\gamma^{\mu}    \left(\frac{\sla{p}-\sla{k}+m}{(p-k)^2-m^2}  \right) 
\gamma^{\nu} \notag \\ &\hspace{6em}\times 
\left( \frac{1}{k^2-\mathcal M ^2(k)} \right)\frac{k_{\mu} k_{\mu}  }{k^2} \,,
\end{align} 
the third one
\begin{align}
\Sigma^{(3)}(p)&= \mu  e^2  \int\frac{\mathrm{d}^3k}{(2\pi)^3}
\gamma^{\mu}    \left(\frac{\sla{p}-\sla{k}+m}{(p-k)^2-m^2}  \right)  
\gamma^{\nu} \notag \\ & \hspace{6em} \times \left(  \frac{  \epsilon_{\mu \beta \nu} 
k^{\beta}}{k^2-\mathcal M ^2(k)}  \right) \frac{ 1 }{k^2} \,,
\end{align} 
and the fourth one
\begin{align}
 \Sigma^{(4)}(p)&=-g  e^2  \int\frac{\mathrm{d}^3k}{(2\pi)^3}
\gamma^{\mu}    \left(\frac{\sla{p}-\sla{k}+m}{(p-k)^2-m^2}  \right)  
\gamma^{\nu} \notag \\ &  \hspace{6em} \times \frac{\epsilon_{\mu \beta \nu} 
k^{\beta} }{k^2-\mathcal M ^2(k)}    \,.
\end{align} 
\end{subequations}
To simplify the calculation of the different pieces, we rewrite 
the pole structure of the gauge propagator into its particle and ghost contributions as
\begin{align} \label{Decompistion_P}
\frac{1}{k^2-\mathcal M^2(k)}&=\frac{1}{g^2 (M_2^2-m_1^2)}  \left(  \frac{1}{
k^2-m_1^2}   \right.  \notag   \\ 
 & \phantom{{}={}} \left.  \hspace{4em} - \frac{1}{k^2-M_2^2}   \right)\,.
\end{align} 
The first term is dominated by the particle mass $m_1$ and the second term by the ghost mass
$M_2$. This
decomposition results to be crucial to radiative corrections.

In order to compute the above corrections we shall use the Feynman parametrization 
\begin{align}
    \frac{1}{AB}=\int_0^1 dx \frac{1}{Ax+B(1-x)}\,.
\end{align}
With suitable choices of $A$ and $B$ together with a shift in 
the momenta, $k+xp\rightarrow l $, and parity properties, the 
first piece of the fermion self-energy in~\eqref{part1} reduces to
\begin{align}
\Sigma^{(1)}(p)&= \frac{ e^2 }{8\pi }    \int  \frac{dx (\sla{p} (1-x)-3m)}{g^2(M_2^2-m_1^2)}
  \notag   \\ 
 & \phantom{{}={}} \times \left[  \frac{1}{   \sqrt{\Delta(m_1)}    
} -\frac{1}{ { \sqrt{ \Delta(M_2)   }    }  }  \right]  \,,
\end{align}
with 
\begin{align}
\Delta(\alpha_i)= p^2x(x-1) +m^2x +\alpha_i^2(1-x)  \,,
\end{align}
where $\alpha_i$ is the notation for either $m_1$ or $M_2$.

In an analogous way, the second piece is
\begin{align}
\Sigma^{(2)}(p)&= \frac{ e^2 }{16\pi }  \int_0^1dx 
\int _0^{1-x}dy   \frac{ 1
 }{g^2(M_2^2-m_1^2)} \\ &
 \left[  3 \alpha  \left ( \frac{1}{   \sqrt{ Q(m_1)} }  - \frac{1}{ \sqrt{ Q(M_2)} }
   \right  )  \right. \nonumber  \\ & \left.  -\beta \left ( \frac{1}{  (Q(m_1))^{3/2}} 
   - \frac{1}{( Q(M_2))^{3/2}}   \right  ) \right]\,,\nonumber  
\end{align}
with 
\begin{align}
\alpha&= \frac{\sla{p}}{3}   (1+5x)+m \,.
\\
\beta&= -\sla{p}p^2(1-x) x^2 +p^2 x^2 m  \,,
\end{align}
\begin{align}
Q(\alpha_i)&= p^2x(x-1)+m^2x+ \alpha_i^2 y \,,
\end{align}
the third piece is
\begin{align}
\Sigma^{(3)}(p)&= -\frac{ e^2 }{16\pi }  \int_0^1dx 
\int _0^{1-x}dy   \frac{ 1
 }{g^2(M_2^2-m_1^2)} \notag \\ &\phantom{{}={}} \times \left[  3 \mu \left ( \frac{1}{ \sqrt{Q(m_1)  } }  -
  \frac{1}{ \sqrt{Q(M_2)  }   }   \right  )  \right. \nonumber   \\   & \left. -\mu \lambda  \left ( \frac{1}{(Q(m_1))^{3/2}}
    - \frac{1}{( Q(M_2))^{3/2}}   \right  ) \right]\,,
\end{align}
and, finally, the fourth piece is
\begin{align} \label{sigma_4}
\Sigma^{(4)}(p)&= \frac{ g e^2 }{2\pi^2 }  \int_0^1dx   \frac{ 1 }
{g^2(M_2^2-m_1^2)} \\ & \left[  - \int _0^{\infty} dr
  \frac{r^4}{(r^2+\Delta)^2}  +\lambda \int _0^{\infty} dr
  \frac{r^2}{(r^2+\Delta)^2} \right]   \nonumber \,,
\end{align}
with 
\begin{align}
\lambda&= m  x \sla{p}-2  p^2(1-x) x  \,.
\end{align}
The first integral in \eqref{sigma_4} has to 
be analytically continued from arbitrary $d$-dimensions to $d=3$ in order to produce a finite result. 

Finally, the total radiative corrections induce in the fermion self-energy the following higher-order operator of the form
\begin{align}
\Sigma(p)&= A(p^2,g,\mu)\sla p -B(p^2,g,\mu)m -gp^2  C(p^2, \mu) \,,
\end{align}
where $A$, $B$ and $C$ are functions that do not contribute to the $p^2$ order. 

Now, focusing on the fermionic part, we  note that the corrections
 induce a general structure in the fermionic Lagrangian of the type
\begin{align}\label{induced_fermion}
\mathcal L_f=\bar \psi (i\sla{\partial}-m)\psi+\bar \psi  g\Box \psi  \,.
\end{align}

We  decompose the above Lagrangian in two parts
\begin{align}
\mathcal L_f=\bar \psi_1 (i\sla{\partial}-\overline m_1)\psi_1-
\bar \psi_2 (i\sla{\partial}-\overline m_2)\psi _2  \,,
\label{fermionlagrangians}
\end{align}
where 
\bea
\psi_1&=& \beta (i\sla \partial -{\overline m}_2 )\psi   \,,
\label{massm1}\\
\psi_2&=& \beta (i\sla \partial - \overline m_1 )\psi   \,,
\label{massm2}
\eea
with 
\begin{align} 
\beta=\left(\frac{g}{\overline m_2-\overline m_1}\right)^{1/2}\,.
\end{align}
The fermionic masses are
\begin{align} \label{fermionmasses1}
\overline{m}_1= &\frac{1-\sqrt{1-4m g}}{2 g} \,,  \\ 
\label{fermionmasses2}
\overline m_2=& \frac{1+\sqrt{1-4 m g}}{2g} \,,
\end{align}
where the first corresponds to a physical fermion 
and the second  to a ghost fermion 
of very high mass of the order $1/g$. It is interesting to note the analogy
of the two fermion species that arises when including 
an odd mass terms in the fermion Lagrangian~\cite{Ayala:2010fm}, which, however are 
physical.

Both propagators can be written as
\begin{align}\label{ferpropa1}
S_1(p)&=\frac{i }  { \sla{p}- \overline m_1}\,,
\\ \label{ferpropa2}
S_2(p)&=\frac{i}{ \sla{p}- \overline m_2}\,.
\end{align}

Note that the fermion propagator described
in~\eqref{induced_fermion} can be also obtained without resorting 
to the explicit decomposition of the fermion fields $\psi_{1,2}$, that is to say, we can 
consider
\begin{align}\label{fermionHD}
 S(p)&\equiv \frac{i}{\slsh{p}-m-gp^2}
  \nonumber \\
  &=\frac{\overline m_1+\overline m_2}{\overline m_2-\overline m_1} \left(S_{1}(p)-S_{2}(p)\right)\,,
\end{align}
with $S_{1,2}(p)$ given in~\eqref{ferpropa1} and~\eqref{ferpropa2}. 
In the next section, we shall use the decomposed expression~\eqref{fermionHD} in order 
to evaluate the photon polarization operator. 
\subsection{Photon polarization operator}\label{subsectionIV-C}
Now we evaluate the photon polarization operator (PPO) 
\begin{align}
    \Pi^{\mu\nu}(q)&=-ie^2\int\frac{\mathrm{d}^3k}{(2\pi)^3}\mathrm{tr}\left[\gamma^\mu S(k) \gamma^\nu S(k+q)
    \right]\,.
\end{align}
For this, we use the decomposed fermion propagator given in~\eqref{fermionHD},
and define the contributions
\begin{align} \label{ppo1}
    \Pi^{\mu\nu}_{ij}(q)=-i\bar{e}^2 \int \frac{\mathrm{d}^3k}{(2\pi)^3}  
     \mathrm{tr} \left[ \gamma^\mu S_i\left(k\right) \gamma^\nu S_j\left(k+q\right)  \right]\,,
\end{align}
with
\begin{align}
S_i(p)&=\frac{i }  { \sla{p}- \overline m_i}\,, \\
  \bar  e &=e \left(\frac{\overline 
  m_1+\overline m_2}{\overline m_2-\overline m_1} \right)\,,
\end{align}
where $i,j =1,2$. We also
introduce the notation for the different contributions
\begin{align}\label{notationPi}
    \Pi^{\mu\nu}(q)
    &\equiv  {\Pi^{\mu\nu}_{11}}(q)-2{\Pi^{\mu\nu}_{12}}(q)+{\Pi^{\mu\nu}_{22}}(q)\,.
\end{align}
Since each contribution in~\eqref{notationPi} has a similar form, we compute the general contribution 
\begin{align}\label{ppof1}
    & \Pi_{ij}^{\mu\nu}(q)=
      2i\bar e^2\int\frac{\mathrm{d}^3k}{(2\pi)^3}\frac{1}
      {(k^2-{\overline m_i}^2)((k+q)^2-{\overline m_j}^2)}\nonumber \\
    &\hspace{1em} \left[k^\mu(k+q)^\nu+
      k^\nu(k+q)^\mu-\eta^{\mu\nu}(k\cdot(k+q)-\overline m_i \overline m_j)\right.
      \nonumber \\
     &\hspace{1em} \left.-i\epsilon^{\mu\nu\beta}(\overline m_i(k+q)_\beta-\overline m_j k_\beta) \right]\,,      
\end{align}
where we have used the expressions~\eqref{diracalgebra} and~\eqref{relationgamma} in the trace 
calculation.

Using the Feynman parametrization together 
with a shift in the momenta, $k+xq\rightarrow l $, and using parity 
properties, the PPO in~\eqref{ppof1} reduces in $d$ dimensions to
\begin{align}
\Pi_{ij}^{\mu\nu}(q)=&2i\bar e^2\mu^{3-d}\int\frac{\mathrm{d}^dl}{(2\pi)^d}
    \int_0^1 dx \frac{1}{[l^2-\Theta_{ij} ]^2}
    \nonumber \\
    & \Bigg[ \left(\frac{2}{d}-1\right)l^2\eta^{\mu\nu} - 2x(1-x)
    (q^\mu q^\nu-\eta^{\mu\nu}q^2) 
    \nonumber \\
    & +\eta^{\mu\nu}(\Theta_{ij} +\overline m_i \overline m_j-(1-x) {\overline m}_i^2-x {\overline m}_j^2) 
    \nonumber \\
    & -i((1-x)\overline m_i+x \overline m_j)q_\beta\epsilon^{\mu\nu\beta} \Bigg]\,,
\end{align}
where we have used dimensional regularization together with the symmetrization 
$l^\mu l^\nu\rightarrow \frac{1}{d}l^2 \eta^{\mu\nu}$ and introduced the element
\begin{align}
\Theta_{ij}   \equiv (1-x)\overline m_i^2+x\overline m_j^2-x(1-x)q^2\,.
\end{align}
In this way, the integral over momenta can be 
straightforwardly performed using the identities given 
in Ref.~\cite{Peskin:1995ev}, arriving at
 \begin{align}\label{ppo2}
 \Pi_{ij}^{\mu\nu}(q)&= \left(\eta^{\mu\nu}-\frac{q^\mu q^\nu}{q^2}\right) {\Pi^e_{ij}}(q)\nonumber \\
 &\hspace{2em}+\eta^{\mu\nu} {\Pi^\eta_{ij}}(q)
 +i\epsilon^{\mu\nu\beta}\frac{q_\beta}{q}{\Pi^o_{ij}}(q)\,,
\end{align}
with
\begin{align}
    {\Pi^e_{ij}}(q)&= - \frac{2{\bar e}^2\mu^{3-d} q^2 \Gamma(2-\frac{d}{2})}{(4\pi)^{d/2}}    
    \int_0^1 dx \frac{2x(1-x)}{\Theta_{ij}^{2-\frac{d}{2}}},\\
{\Pi^\eta_{ij}}(q)&= - \frac{2\bar e^2\mu^{3-d} \Gamma(2-\frac{d}{2}) }{(4\pi)^{d/2}}\nonumber \\
&\hspace{2em}\times \int_0^1 dx \frac{\overline m_i \overline m_j-(1-x){\overline m}_i^2
-x{\overline m}_j^2}{\Theta_{ij}^{2-\frac{d}{2}}}\,,
\\
{\Pi^o_{ij}}(q)&= \frac{2{\bar e}^2\mu^{3-d}q \Gamma(2-\frac{d}{2})}{(4\pi)^{d/2}}
\int_0^1 dx \frac{(1-x)\overline m_i+x\overline m_j}{\Theta_{ij}^{2-\frac{d}{2}}}.
\end{align}
To analyze the divergent behavior in the PPO in $2+1$ dimensions, let us
introduce the dimensional regularization
requiring the space-time dimension to
be $d=3-\epsilon$ and take the limit $\epsilon\rightarrow0$, obtaining 
\begin{align}
{\Pi^e_{ij}}(q)&= -\frac{4{\bar e}^2 q^2\sqrt{\pi} }{(4\pi)^{3/2}}\int_0^1 
dx \frac{x(1-x)}{\Theta_{ij}^\frac{1}{2}}\,,\\
{\Pi^\eta_{ij}}(q)&= -\frac{2{\bar e}^2 \sqrt{\pi}}{(4\pi)^{3/2}}\int_0^1 
dx \frac{\overline m_i \overline m_j-(1-x){\overline m}_i^2-x{\overline m}_j^2}
{\Theta_{ij}^{\frac{1}{2}}}\,,
\\
{\Pi^o_{ij}}(q)&= \frac{2{\bar e}^2 q\sqrt{\pi}}{(4\pi)^{3/2}}\int_0^1 dx
\frac{(1-x)\overline m_i+x\overline m_j}{\Theta_{ij}^{\frac{1}{2}}}\,.
\end{align}
All these contributions are finite, as can be seen 
from the absence of $1/\epsilon$ terms.

After integrating over the Feynman parameter, we have
\begin{align}
&{\Pi^e_{ij}}(q)=\nonumber \\
&\hspace{2em} -\frac{4{\bar e}^2\sqrt{\pi} q^2}{(4\pi)^{3/2}}\left[\frac{3({\overline m}_i^2
-{\overline m}_j^2)^2-2({\overline m}_i^2+\overline{m}_j^2)q^2-q^4}{8 q^5} \right. 
\nonumber \\
&\times \left.
 \ln\left|\frac{ \overline m_i+\overline m_j-q}{\overline m_i
 +\overline m_j+q}\right|\right. 
\left.+\frac{(\overline m_i+\overline m_j)(3(\overline m_i
-\overline m_j)^2-q^2)}{4 q^4}\right]\,,
\end{align}
\begin{align}
     {\Pi^\eta_{ij}}(q)&= 2{\bar e}^2\frac{\pi^{1/2}}{(4\pi)^{3/2}}\frac{(\overline 
     m_i+\overline m_j)(\overline m_i-\overline m_j)^2}{q^2}  \\
 &\times \left[1+\frac{(\overline m_i+\overline m_j)^2-q^2}{2q(\overline m_i+\overline m_j)}
 \ln\left|\frac{\overline m_i+\overline m_j-q}{\overline m_i+\overline m_j+q}\right|\right]\,, \nonumber 
\end{align}
and
\begin{align}
  {\Pi^o_{ij}}(q)&= 2{\bar e}^2\frac{\pi^{1/2}}{(4\pi)^{3/2}}q
 \left[\frac{(\overline m_i-\overline m_j)^2}{q^2}\right. \\
 &\left.+\frac{(\overline m_i+\overline m_j)((\overline m_i-\overline m_j)^2-q^2)}{2q^3}
 \ln\left|\frac{ \overline m_i+\overline m_j-q}{\overline m_i+\overline m_j+q}\right|\right]  \,. \nonumber 
\end{align}
Taking into account the above-defined notation, 
in an analogous way, we define for the pieces $e,o,\eta$ the total contributions
$\Pi^{e,\eta,o}=  \Pi^{e,\eta,o}_{11}- 
2\Pi^{e,\eta,o}_{12}+ \Pi^{e,\eta,o}_{22} $, and we can see that
\begin{align}\label{ppoHD}
&\Pi^e(q)=-4{\bar e}^2\frac{\pi^{1/2}}{(4\pi)^{3/2}}\frac{q^2(\overline m_1+
\overline m_2)^2}{(\overline m_2-\overline m_1)^2}
 \\
&\hspace{.5cm} 
 \left\{\frac{-4\overline m_1^2-q^2}{8 q^3} \ln\left|\frac{ 2\overline m_1-q}{2\overline m_1+q}\right|
\right.  \nonumber  \\ & \left.  -\frac{2\overline m_1}{4 q^2}+\frac{-4\overline m_2^2-q^2}{8 q^3} 
\ln\left|\frac{ 2\overline m_2-q}{2\overline m_2+q}\right|-\frac{2\overline m_2}{4 q^2}\right.
\nonumber \\
&\hspace{.5cm} \left.-2\frac{3({\overline m}_1^2-{\overline m}_2^2)^2-2({\overline m}_1^2+{\overline m}_2^2)q^2-q^4}{8 q^5} \right. 
\nonumber \\   & \times \left.  \ln\left|\frac{ \overline m_1+\overline m_2-q}{\overline m_1+\overline m_2+q}\right|
-2\frac{(\overline m_1+\overline m_2)(3(\overline m_1-\overline m_2)^2-q^2)}{4 q^4}\right\}  \nonumber \,,                                 
\end{align}
\begin{align}\label{ppoHD2}
\Pi^\eta(q)&= 2{\bar e}^2\frac{\pi^{1/2}}{(4\pi)^{3/2}}\frac{(-2)(\overline m_1+\overline m_2)(\overline m_1-\overline m_2)^2}{q^2}
   \notag  \\ &   \left\{1+\frac{(\overline m_1+\overline m_2)^2-q^2}{2q(\overline m_1
   +\overline m_2)}\ln\left|\frac{\overline m_1+\overline m_2-q}{\overline m_1+\overline m_2+q}\right|\right\}
\,,
\end{align}
and
\begin{align}\label{ppoHD3}
\Pi^o(q)&= 2{\bar e}^2\frac{\pi^{1/2}}{(4\pi)^{3/2}}\frac{q(\overline m_1+\overline m_2)^2}{(\overline m_1-\overline m_2)^2}
\notag  \\ &   \left\{-\frac{\overline m_1}{q}\ln\left|\frac{ 2\overline m_1-q}{2\overline m_1+q}\right|
 -\frac{\overline m_2}{q}\ln\left|\frac{ 2\overline m_2-q}{2\overline m_2+q}\right|\right.
 \nonumber \\
&\left.-2\frac{(\overline m_1-\overline m_2)^2}{q^2}-2\frac{(\overline m_1+\overline m_2)((\overline m_1-\overline m_2)^2-q^2)}{2q^3}
\right. \notag  \\ &\times   \left. \ln\left|\frac{ \overline m_1+\overline m_2-q}{\overline m_1+\overline m_2+q}\right|\right\}\,.
\end{align}

Let us explore the behavior of the contributions~\eqref{ppoHD}, \eqref{ppoHD2} and \eqref{ppoHD3}, when $g$
is very small. In this case the mass $\overline m_2$ becomes the dominant energy scale 
compared to $\overline m_1\rightarrow m$, which yields  
\begin{align}
&\Pi^e(q)  {\approx}-4e^2\frac{\pi^{1/2}}{(4\pi)^{3/2}}q^2
\left\{-\frac{4m^2+q^2}{8 q^3}
\left(1+\frac{4m}{\overline m_2}\right) \right.    \notag  \\& 
\left.\hspace{2em}
\times \ln\left|\frac{ 2m-q}{2m+q}\right| -\frac{2m}{4q^2}
  -\frac{11}{30 \overline m_2}-\frac{2m^2}{q^2\overline m_2} 
\right\}\,,                                    
\end{align}
\begin{align}
   &\Pi^\eta(q)     {\approx}2e^2\frac{\pi^{1/2}}{(4\pi)^{3/2}}
\left\{\frac{4}{3}\left(3m-\overline m_2-\frac{4m^2}{\overline m_2}\right) - \frac{4 q^2}{15 \overline m_2}\right\}\,, 
\end{align}
and 
\begin{align}
\Pi^o(q)     &{\approx}2e^2\frac{\pi^{1/2}}{(4\pi)^{3/2}}q
\left\{-\frac{m}{q}\left(1+\frac{4m}{\overline m_2}\right)\ln\left|\frac{ 2m-q}{2m+q}\right|
  \right. \notag \\& \left. -\frac{1}{3} - \frac{4 m}{\overline m_2}\right\}\,.
\end{align}

Another interesting scenario corresponds to the
UV limit so that the higher derivative terms 
have important contributions. In this case with $q\rightarrow \infty$, each coefficient has the form  
\bea
\Pi^e(q)     &\stackrel{q\rightarrow \infty}{\approx}&-4{\bar e}^2\frac{\pi^{1/2}}{(4\pi)^{3/2}}q^2
\left\{-i\pi\frac{3(\overline m_1+\overline m_2)^4}{4q^5}\right.  \\&& \left. +\frac{16(\overline m_1
+\overline m_2)^3({\overline m}_1^2+{\overline m}_1\overline m_2+{\overline m}_2^2)}{3q^6}\right\}. \nonumber 
\eea
Note that $\Pi^e$ in our case
has a better ultraviolet behavior 
than in the case of the absence of
 higher derivative contributions where it has the form
\begin{align}
 \Pi^e(q)    &\stackrel{q\rightarrow \infty}{\approx}& -4e^2\frac{\pi^{1/2}}{(4\pi)^{3/2}}q^2
 \left\{ -i\pi\frac{q^2+4m^2}{8q^3}+\frac{8m^3}{3q^4}\right\}.    
\end{align}

For the metric coefficient, we have
\begin{align}
\Pi^\eta(q)     &\stackrel{q\rightarrow \infty}{\approx}&2{\bar e}^2\frac{\pi^{1/2}}{(4\pi)^{3/2}}
\left\{ i\pi \frac{(\overline m_1 +\overline m_2)^2 }{q} -\frac{4  (\overline m_1 + \overline m_2)^3}{q^2} \right\}.
\end{align}
Note that in the absence of higher derivative contributions, this term does not exist.

Finally, in this limit the odd term gives
\bea
 \Pi^o(q)     &\stackrel{q\rightarrow \infty}{\approx}&2{\bar e}^2\frac{\pi^{1/2}}{(4\pi)^{3/2}}q
\left\{-i\pi\frac{(\overline m_1+\overline m_2)^3}{q^3}\right.  \\&& \left.\notag +\frac{4(\overline 
m_1+\overline m_2)^2(5{\overline m}_1^2+8\overline m_1 \overline m_2+5{\overline m}_2^2)}{3q^4}\right\}\,,
\eea 
from where we notice that the ultraviolet behavior is different to the 
standard CS QED result (see e.g. \cite{Chaichian:1997ix}):
\bea
 \Pi^o(q)     &\stackrel{q\rightarrow \infty}{\approx}&2{\bar e}^2\frac{\pi^{1/2}}{(4\pi)^{3/2}}q
\left\{-i\pi\frac{m}{q}+\frac{4m^2}{q^2}\right\}.   
\eea
\subsection{The vertex diagram}\label{subsectionIV-D}
Let us expand the propagator
$S(k+p)$ given in \eqref{fermionHD}, into power series 
in the external momentum $p$ up to  the first order as 
\begin{align}
S(k+p)=S(k)-S(k)\sla{p}S(k)+\ldots   \,, 
\end{align}
in such a way that the two-point function of the 
spinor field given in~\eqref{fermionselfP} can be expanded as
\begin{align}
    \Sigma(p)=\Sigma(0)+\Sigma_1(p)+\ldots
\end{align}
with
\begin{align}\label{2point}
\Sigma_1(p)=-e^2\int\frac{d^3k}{(2\pi)^3}\gamma^{\alpha}S(k)
\sla{p}S(k)\gamma^{\beta}\Delta_{\alpha\beta}(k)+\ldots  \,,
\end{align}
where dots are for irrelevant terms of zero, second, and higher orders.

On the other hand, the three-point vertex function, in zero order
in external momenta and in the Landau gauge, looks like
\begin{align}
\Gamma^{\mu}(0,0)=-e^2\int\frac{d^3k}{(2\pi)^3}\gamma^{\alpha}S(k)
\gamma^{\mu}S(k)\gamma^{\beta}\Delta_{\alpha\beta}(k).
\end{align}
Comparing this expression with~\eqref{2point}, we see that they 
are equal up to the overall factor. 
Therefore we see that the corrections to the kinetic term for 
the spinor field and the triple spinor-vector 
vertex are described by the same integral over momenta, hence, the gauge symmetry is not jeopardized 
by quantum corrections which is consistent with the Ward identities up to first 
order in external momenta. In other words, the "covariant derivative" 
term $\bar{\psi}\sla{D}\psi$ receives the quantum correction $ \bar{\psi}\Gamma^{\mu}D_{\mu}\psi$, where
\begin{align}
\Gamma^{\mu}(0,0)=-e^2\int\frac{d^3k}{(2\pi)^3}\gamma^{\alpha}S(k)
\gamma^{\mu}S(k)\gamma^{\beta}\Delta_{\alpha\beta}(k).
\end{align}

Now it is instructive to check the gauge invariance of the quantum correction generated on the 
 spinor sector which as it is well known, is deeply related with the Ward identities. 
 We proceed with the computation of the vertex function
 and take advantage of the previous computation.
 
Within our study, we consider the lower (zero) order contributions to the
 three-point function only. Therefore, we require $p=0$ and $p'=0$ and arrive at
\begin{align}
\Gamma^{\mu}(0,0)=-e^2\int\frac{\mathrm{d}^3k}{(2\pi)^3}\gamma^{\alpha}S(k)
\gamma^{\mu}S(k)\gamma^{\beta}\Delta_{\alpha\beta}(k).
\label{vertex1}
\end{align}
 
Similarly to the above PPO calculation, we use the decomposed 
fermion propagator~\eqref{fermionHD} in \eqref{vertex1}, so, the vertex contributions read
\begin{align}\label{vertexgen}
\Gamma_{ij}^{\mu}(0,0)=-{\bar e}^2\int\frac{\mathrm{d}^3k}{(2\pi)^3}\gamma^{\alpha}S_i(k)
\gamma^{\mu}S_j(k)\gamma^{\beta}\Delta_{\alpha\beta}(k).
\end{align}
where $i,j=1,2.$ As before, we introduce the notation for the different contributions
\begin{align}
\Gamma^{\mu}(0,0)=\Gamma_{11}^\mu(0,0)-2\Gamma_{12}^\mu(0,0)+\Gamma_{22}^\mu(0,0).
\end{align}

Since all contributions have the same form, we calculate its general form given by
\begin{align} \label{vertex2}
     \Gamma_{ij}^\mu(0,0)
          &=-i {\bar e}^2   \int\frac{\mathrm{d}^3k}{(2\pi)^3}
     \frac{\gamma^\rho (\slsh{k}-\overline{m}_i)\gamma^\mu(\slsh{k}-\overline m_j)\gamma^\sigma}
            {(k^2-{\overline m}_i^2)(k^2-{\overline m}_j^2)}\nonumber \\ 
            &
        \times \frac{\gamma T_{\rho\sigma}(k)}{k^2-\gamma^2\mathcal{M}^2}\,.    
\end{align}
Using the identity 
\begin{align}
&(\slsh{k}-\overline{m}_i)\gamma^\mu(\slsh{k}-\overline m_j)=\nonumber \\
&\hspace{1em}-k^2\gamma^\mu +2\sla{k}k^\mu -\sla{k}\gamma^\mu \overline m_j -\overline m_i 
\gamma^\mu \sla{k}+ \overline m_i\overline m_j \gamma^\mu\,,
\end{align}
together with \eqref{relationgamma}, the three-point vertex function in~\eqref{vertex2}  becomes
\begin{align}
\Gamma_{ij}^\mu(0,0)&=   ({\Gamma_{ij}^\mu})^a+  ({\Gamma_{ij}^\mu})^b,
\end{align}
where the $(\Gamma^{\mu}_{ij})^a$ and $(\Gamma^{\mu}_{ij})^b$ are the contributions without 
and with the Levi-Civita symbol coming from the gauge boson propagator, respectively. These are given by
\begin{align}
({\Gamma_{ij}^\mu})^a
&=-i{\bar e}^2   \int\frac{\mathrm{d}^3k}{(2\pi)^3}
\frac{\gamma^\rho (-k^2\gamma^\mu +2\sla{k}k^\mu+ \overline m_i\overline m_j \gamma^\mu)
\gamma^\sigma } {(k^2-{\overline m}_i^2)(k^2-{\overline m}_j^2)} \nonumber\\
&\hspace{2em}\times \frac{  \gamma(  \eta_{  \rho \sigma }-  
 k_{\rho}k_{\sigma}/k^2 ) }{k^2-\gamma^2\mathcal{M}^2},      
\end{align}
and 
\begin{align}
({\Gamma_{ij}^\mu})^b
&={\bar e}^2   \int\frac{\mathrm{d}^3k}{(2\pi)^3}
\frac{\gamma^\rho ( \sla{k}\gamma^\mu \overline m_j +\overline m_i
\gamma^\mu \sla{k})\gamma^\sigma } {(k^2-{\overline m}_i^2)(k^2-{\overline m}_j^2)} \nonumber\\
&\hspace{2em}\times \frac{  \gamma( \gamma \mathcal{M}\,\epsilon_{\rho\beta\sigma}k^\beta /k^2) }{k^2-\gamma^2\mathcal{M}^2}\,.     
\end{align}
Now,  we use~\eqref{relationgamma} together with symmetry 
properties on the integral over momenta in the above equations, getting
\begin{align}
({\Gamma_{ij}^\mu})^a&=i{\bar e}^2\gamma \gamma^{\mu}
 \\
&\hspace{1em} \times 
\int\frac{\mathrm{d}^3k}{(2\pi)^3}\frac{k^2+\frac{1}{3}\overline m_i
\overline m_j}{(k^2-{\overline m}_i^2)(k^2-{\overline m}_j^2)(k^2-\gamma^2\mathcal{M}^2)}.
\nonumber
\end{align}
and 
\begin{align}
    ( {\Gamma_{ij}^\mu})^b&=i{\bar e}^2 \gamma
    \gamma^\mu
          \nonumber\\
     &\times \int\frac{d^3k}{(2\pi)^3}
     \frac{\frac{10}{3}\gamma(\overline m_i+\overline m_j)\mathcal{M} }
            {(k^2-{\overline m}_i^2)(k^2-{\overline m}_j^2)(k^2-\gamma^2\mathcal{M}^2)}    \,.
\end{align}

Once we carried out the Wick rotation, we have two integrals of the form 
\begin{align}
I_{1}&=\int_0^\infty
     \frac{r^2 dr}
            {(   { \gamma^2}g^2 r^4+{\gamma^2} 2 g \mu  r^2+{ \gamma^2}\mu ^2+r^2)}\nonumber \\
      &\hspace{1em} \times \frac{1}{(r^2+{\overline m}_i^2)(r^2+{\overline m}_j^2) }     \,,
\end{align}
and
\begin{align}
I_{2}&=\int_0^\infty
     \frac{r^4 dr}
            {({ \gamma^2}g^2 r^4+{\gamma^2} 2 g \mu  r^2+{ \gamma^2}\mu ^2+r^2)}\nonumber \\
      &\hspace{1em} \times \frac{1}{(r^2+{\overline m}_i^2)(r^2+{\overline m}_j^2) }     \,,
\end{align}
which can be easily performed using a standard procedure, yielding
\begin{align}
I_{1}&=-\frac{\pi}{2g^2\gamma^2}\frac{m_1+M_2+\overline m_i +\overline m_j}
{(m_1+\overline m_i)(m_1+\overline m_j)(M_2+\overline m_i)(M_2+\overline m_j)}
\nonumber \\
    &\hspace{1em}\times \frac{1}{(\overline m_i+\overline m_j)(m_1+M_2)} \,,
\end{align}
and 
\begin{align}
I_{2}&=-\frac{\pi}{2g^2\gamma^2}\frac{(m_1+M_2)(\overline m_i \overline m_j)
+(\overline m_i +\overline m_j)(m_1M_2)}{(m_1+\overline m_i)(m_1+\overline m_j)(M_2+\overline m_i)(M_2+\overline m_j)}
\nonumber \\
    &\hspace{1em}\times \frac{1}{(\overline m_i+\overline m_j)(m_1+M_2)}.
\end{align}

With these results, we can see that $\Gamma^\mu= 
\Gamma^{\mu}_{11}-2\Gamma^{\mu}_{12}+ \Gamma^{\mu}_{22} $ has the form
\begin{align}
   &\Gamma^\mu=\frac{\bar e^2 }{4\pi g^2\gamma}\left\{\frac{\left(2 \overline{m}_1+m_1+M_2\right) 
   \left({\overline m}_1+20\gamma \mu\right)}{6
   \left(m_1+M_2\right)
   \left({\overline m}_1+m_1\right){}^2
   \left({\overline m}_1+M_2\right){}^2}\right.
\nonumber \\
&\hspace{0em}-\frac{\left({\overline m}_1
   \left(m_1+M_2\right)+2 m_1 M_2\right)
   \left(3+20 \gamma  g {\overline m}_1\right)}{6
   \left(m_1+M_2\right)
   \left({\overline m}_1+m_1\right){}^2
   \left({\overline m}_1+M_2\right){}^2}
   \nonumber\\
    &\hspace{0em}+
    \frac{\left(2 {\overline m}_2+m_1+M_2\right) 
    \left({\overline m}_2+20\gamma  \mu \right)}{6
   \left(m_1+M_2\right)
   \left({\overline m}_2+m_1\right){}^2
   \left({\overline m}_2+M_2\right){}^2}
   \nonumber\\
   &\hspace{0em} -\frac{\left({\overline m}_2
   \left(m_1+M_2\right)+2 m_1 M_2\right)
   \left(3+20 \gamma  g {\overline m}_2\right)}{6
   \left(m_1+M_2\right)
   \left({\overline m}_2+m_1\right){}^2
   \left({\overline m}_2+M_2\right){}^2}
   \nonumber\\  
   &\hspace{0em}-\frac{2}{
   \left({\overline m}_1+{\overline m}_2\right) \left(m_1+M_2\right)
   \left({\overline m}_1+m_1\right)}
   \nonumber\\  
   &\hspace{0em}\times
   \left[\frac{\left({\overline m}_1+{\overline m}_2+m_1+M_2\right)
   \left(10 \gamma  \mu 
   \left({\overline m}_1+{\overline m}_2\right)+{\overline m}_1 {\overline m}_2\right)}{3\left({\overline m}_2+m_1\right)
   \left({\overline m}_1+M_2\right) \left({\overline m}_2+M_2\right)}
   \right.
   \nonumber \\  
   &\hspace{1em}+\frac{
   m_1 M_2({\overline m}_1 +{\overline m}_2)
   \left(1+\frac{10}{3} \gamma  g
   \left({\overline m}_1+{\overline m}_2\right)\right)}{
    \left({\overline m}_2+m_1\right)
   \left({\overline m}_1+M_2\right) \left({\overline m}_2+M_2\right)}
   \nonumber \\     
   &\hspace{1em}\left.\left.+\frac{
   {\overline m}_1 {\overline m}_2 
   \left(m_1+M_2\right)
   \left(1+\frac{10}{3} \gamma  g
   \left({\overline m}_1+{\overline m}_2\right)\right)}{
    \left({\overline m}_2+m_1\right)
   \left({\overline m}_1+M_2\right) \left({\overline m}_2+M_2\right)}
   \right]\right\}\gamma^{\mu}\,.
   \label{vertexresultF1}
\end{align}
Since this one-loop correction to the vertex has a simple $\gamma^\mu$ 
structure, its coefficient can be identified as the form factor $F_1(q^2=0)$, so the vertex can be written as
\begin{align}
   &\Gamma^\mu=F_1(0)\gamma^\mu\,,
\end{align}
which is expected, as we set $p=p'=0$, eliminating any other available tensor structure.

As \eqref{vertexresultF1} is a complete result for an arbitrary value of $\gamma$,  
to make contact with the literature, let us consider a pure CS theory  
by taking the limit $\gamma\rightarrow\infty$ in~\eqref{vertexresultF1}, getting 
\begin{align}
    &F_1(0)\stackrel{\gamma\rightarrow\infty}{=}\frac{e^2}{4\pi}\frac{5 \left(\sqrt{\mu }-\sqrt{g} (m-3 \mu
   )\right)}{3 \left(\sqrt{g} (\mu +m)+\sqrt{\mu
   }\right)^3}\,,
   \label{F1PCS}
\end{align}
where we use~\eqref{bosonmasses1},~\eqref{bosonmasses2},~\eqref{fermionmasses1} 
and~\eqref{fermionmasses2}. Additionally, if we take off the higher derivative 
contributions in~\eqref{F1PCS} by setting~$g\rightarrow0$, we arrive at~\cite{Chen}
\begin{align}
    &F_1(0)\stackrel{g\rightarrow0}{=}\frac{e^2}{4\pi}\frac{5}{3\mu}.
\end{align}
\section{Microcausality in the gauge sector}\label{sectionV}
The causal structure of a quantum field theory is determined by the behavior of the commutator of fundamental 
fields at two distinct spacetime points, $x$ and $y$, outside the light cone. In what follows, we analyze 
the implications of this criterion for causality in the gauge sector.
Hence, let us consider the commutator   
\begin{align}\label{causint}
D_{\mu\nu}(x-y)=\left[A_{\mu}(x),A_{\nu}(y)\right]\,,
\end{align}
and verify whether it vanishes outside 
the lightcone, i.e., when $(x-y)^2<0$. 

From the decomposition of the 
gauge field~\eqref{gauge_decomposition}, we have the
contributions
\begin{align}
    D_{\mu\nu}(x-y)&=\big[\bar{A}_\mu(x),\bar{A}_\nu(y)\big]+\big[G_\mu(x),G_\nu(y)\big]\,.
\end{align}
The first contribution is
\begin{align}\label{commutator-BarABarA}
 & \big[\bar{A}_\mu(x),\bar{A}_\nu(y)\big]=-\int \frac{\mathrm{d}^2\vec{k}}{(2\pi)^2}\frac{1}
 {\Lambda_-'(\omega_1,\vec{k})}\bigg\lbrace \varepsilon_\mu^{(-)}(k_0,\vec{k}) \notag \\
 & \hspace{6em} \times \varepsilon_\nu^{(-)*}(k_0,\vec{k})
  e^{-i k\cdot (x-y)} \notag \\ & \hspace{2em}-\varepsilon_\mu^{(-)*}(k_0,\vec{k}) 
  \varepsilon_\nu^{(-)}(k_0,\vec{k})e^{i k\cdot (x-y)} \bigg\rbrace_{k_0=\omega_1} \,,
\end{align}
and the second
\begin{align}\label{commutator-GG}
 &  \big[G_\mu(x),G_\nu(y)\big]= -\int \frac{\mathrm{d}^2\vec{k}}{(2\pi)^2}\frac{1}
   {\Lambda_+'(W_2,\vec{k})}\bigg\lbrace
   \varepsilon_\mu^{(+)}(k_0,\vec{k})\notag \\ & \hspace{6em}\times
   \varepsilon_\nu^{(+)*}(k_0,\vec{k})   e^{-i k\cdot (x-y)}
   \notag \\ & \hspace{2em}-\varepsilon_\mu^{(+)*}(k_0,\vec{k}) 
   \varepsilon_\nu^{(+)}(k_0,\vec{k})e^{i k\cdot (x-y)} \bigg\rbrace_{k_0=W_2} \,.
\end{align}
Let us define $z=x-y$ and change 
$\vec{k}\rightarrow -\vec{k}$ in the second 
terms of Eqs~\eqref{commutator-BarABarA} and~\eqref{commutator-GG}, obtaining
\begin{align}
 & \big[\bar{A}_\mu(x),\bar{A}_\nu(y)\big]=-\int 
  \frac{\mathrm{d}^2\vec{k}}{(2\pi)^2}\frac{e^{i\vec{k}\cdot\vec{z}}}
  {\Lambda_-'(\omega_1,\vec{k})}\bigg\lbrace \varepsilon_\mu^{(-)}(\omega_1,\vec{k})
  \notag \\ & \hspace{6em}\times  \varepsilon_\nu^{(-)*}(\omega_1,\vec{k})e^{-i \omega_1 z_0}
  \notag \\ &  \hspace{2em} -\varepsilon_\mu^{(-)*}(\omega_1,-\vec{k}) 
  \varepsilon_\nu^{(-)}(\omega_1,-\vec{k})e^{i \omega_1 z_0}\bigg\rbrace  \,,
  \end{align}
  and
  \begin{align}
 & \big[G_\mu(x),G_\nu(y)\big]=-\int 
 \frac{\mathrm{d}^2\vec{k}}{(2\pi)^2}\frac{e^{i\vec{k}\cdot\vec{z}}}
 {\Lambda_+'(W_2,\vec{k})}\bigg\lbrace \varepsilon_\mu^{(+)}(W_2,\vec{k}) \notag
 \\ & \hspace{6em}\times \varepsilon_\nu^{(+)*}(W_2,\vec{k})e^{-i W_2 z_0}
 \notag \\ & \hspace{2em}-\varepsilon_\mu^{(+)*}(W_2,-\vec{k})
 \varepsilon_\nu^{(+)}(W_2,-\vec{k})e^{i W_2 z_0} \bigg\rbrace  \,,
\end{align}
where we have used that 
the denominators are even functions of $\vec k$, see the relations~\eqref{lambdas}
and \eqref{lambdas2}.

We recall the identity \eqref{relationspol}
\begin{align}
   & \varepsilon_\mu^{(\pm)}(k_0,\vec{k})   \varepsilon_\nu^{(\pm)*}(k_0,\vec{k})  
    =
    -\frac{1}{2}\bigg(\eta_{\mu\nu}-\frac{k_\mu k_\nu}{k^2} \notag \\ & \hspace{4em} \pm
    i\epsilon_{\mu\beta\nu}\frac{k^\beta}{\sqrt{k^2}}\bigg) \,.
\end{align}
We take the complex conjugate
\begin{align}
   \varepsilon_\mu^{(\pm)*}(k_0,\vec{k})  \varepsilon_\nu^{(\pm)}(k_0,\vec{k})&=
   -\frac{1}{2}\bigg(\eta_{\mu\nu}  -
   \frac{k_\mu k_\nu}{k^2} \notag \\ & \hspace{1em} \mp i\epsilon_{\mu\beta\nu}\frac{k^\beta}{\sqrt{k^2}}\bigg)\,.
\end{align}
Thus, we conclude that
\begin{align}
     \varepsilon_\mu^{(\pm)*}(k_0,\vec{k})  \varepsilon_\nu^{(\pm)}
     (k_0,\vec{k})& = \varepsilon_\mu^{(\pm)}(-k_0,-\vec{k}) 
     \notag \\  & \hspace{1em} \times \varepsilon_\nu^{(\pm)*}(-k_0,-\vec{k})  \,,
\end{align}
which leads to the relations
\begin{align}
    \varepsilon_\mu^{(-)*}(\omega_1,-\vec{k})  \varepsilon_\nu^{(-)}(\omega_1,-\vec{k})
    &=\varepsilon_\mu^{(-)}(-\omega_1,\vec{k}) \notag \\ &\hspace{1em} \times   \varepsilon_\nu^{(-)*}(-\omega_1,\vec{k}) \,,
   \\
    \varepsilon_\mu^{(+)*}(W_2,-\vec{k})  \varepsilon_\nu^{(+)}(W_2,-\vec{k})
    &= \varepsilon_\mu^{(+)}(-W_2,\vec{k})  \notag \\ & \hspace{1em} \times   \varepsilon_\nu^{(+)*}(-W_2,\vec{k}) \,.
\end{align}
Using these results, the total expression for the commutator is
\begin{align}
    D_{\mu\nu}(z)&=-\int \frac{\mathrm{d}^2\vec{k}}{(2\pi)^2}\frac{e^{i\vec{k}\cdot\vec{z}}}
    {\Lambda_-'(\omega_1,\vec{k})}\bigg\lbrace \varepsilon_\mu^{(-)}(\omega_1,\vec{k})
    \varepsilon_\nu^{(-)*}(\omega_1,\vec{k})   \notag \\ &   e^{-i \omega_1 z_0}
    -\varepsilon_\mu^{(-)}(-\omega_1,\vec{k})  \varepsilon_\nu^{(-)*}
    (-\omega_1,\vec{k})e^{-i(- \omega_1) z_0}\bigg\rbrace \notag \\
    & -\int \frac{\mathrm{d}^2\vec{k}}{(2\pi)^2}\frac{e^{i\vec{k}\cdot\vec{z}}}
    {\Lambda_+'(W_2,\vec{k})}\bigg\lbrace \varepsilon_\mu^{(+)}(W_2,\vec{k})\varepsilon_\nu^{(+)*}(W_2,\vec{k})
    \notag \\ &  e^{-i W_2 z_0} -\varepsilon_\mu^{(+)}(-W_2,\vec{k})  \varepsilon_\nu^{(+)*}(-W_2,\vec{k})e^{-i(- W_2) z_0} \bigg\rbrace  \,.
\end{align}
We consider
\begin{eqnarray}
    \Lambda_-'(k)&=&2k_0\bigg(-\frac{1}{\gamma}
    +\frac{\mathcal{M}(k)}{2\sqrt{k^2}} -g\sqrt{k^2} \bigg)  \,,
    \\
    \Lambda_+'(k)&=&2k_0\bigg(-\frac{1}{\gamma}-\frac{\mathcal{M}(k)}{2\sqrt{k^2}} +g\sqrt{k^2} \bigg) \,,
\end{eqnarray}
and by evaluating in their respective frequencies we recall
what we have obtained before
\begin{eqnarray}
    \Lambda_-'(k)\vert_{k_0=\pm \omega_1}&=&\mp 2\omega_1
    \bigg(\frac{\sqrt{1+4\gamma^2\mu g}}{2\gamma}\bigg) \,, \\
    \Lambda_+'(k)\vert_{k_0=\pm W_2}&=&\pm 2 W_2\bigg(\frac{\sqrt{1+4\gamma^2\mu g}}{2\gamma}\bigg)  \,.
\end{eqnarray}
By the other hand we recall some previous results~\eqref{Tmunu},
and by evaluating $k_0$ in a root of $k^2-\gamma^2\mathcal{M}^2$, we arrive at
\begin{eqnarray}
    T_{\mu\nu}(p)\vert_{\Lambda_\pm=0}=\eta_{\mu\nu}
    -\frac{k_\mu k_\nu}{k^2}\pm i\epsilon_{\mu\beta\nu}\frac{k^\beta}{\sqrt{k^2}} \,.
\end{eqnarray}
We can establish a connection with the polarization vectors
\begin{align}
    \varepsilon_\mu^{(-)}(\pm\omega_1,\vec{k})\varepsilon_\nu^{(-)*}
    (\pm\omega_1,\vec{k})   &=-\frac{1}{2}T_{\mu\nu}(k)\vert_{k_0=\pm\omega_1} \,,
\\
    \varepsilon_\mu^{(+)}(\pm W_2,\vec{k})\varepsilon_\nu^{(+)*}(\pm W_2,\vec{k}) 
    &=-\frac{1}{2}T_{\mu\nu}(k)\vert_{k_0=\pm W_2} \,.
\end{align}
Therefore
\begin{align}
    D_{\mu\nu}(z)&=-\frac{\gamma}{\sqrt{1+4\gamma^2\mu g}}\int \frac{\mathrm{d}^2\vec{k}}{(2\pi)^2}e^{i\vec{k}
    \cdot\vec{z}}\bigg\lbrace \frac{T_{\mu\nu}(k)\vert_{k_0=\omega_1}}{ 2\omega_1} \notag
    \\ & \times e^{-i \omega_1 z_0}   +\frac{T_{\mu\nu}(k)\vert_{k_0=-\omega_1}}{ 2(-\omega_1)}
    e^{-i(- \omega_1) z_0}\bigg\rbrace \notag \\
    &  +\frac{\gamma}{\sqrt{1+4\gamma^2\mu g}}\int \frac{\mathrm{d}^2
    \vec{k}}{(2\pi)^2}e^{i\vec{k}\cdot\vec{z}}
    \bigg\lbrace \frac{T_{\mu\nu}(k)\vert_{k_0=W_2}}{2 W_2 }
    \notag \\ & \times e^{-i W_2 z_0} +\frac{T_{\mu\nu}(k)\vert_{k_0=- W_2}}{2(- W_2) }e^{-i(- W_2) z_0} \bigg\rbrace  \,.
\end{align}
We introduce the contour integral $\mathcal C$ in the 
complex $k_0$-plane that encloses all four poles in the 
counterclockwise direction, and write
\begin{align}
D_{\mu\nu}(z)&=-\frac{\gamma}{\sqrt{1+4\gamma^2\mu g}}\int 
\frac{\mathrm{d}^2\vec {k}}{(2\pi)^2}\oint_\mathcal{C}\frac{\mathrm{d}k_0}{(2\pi i)}
    e^{-i k\cdot z}   \\   &   \times T_{\mu\nu}(k)    \bigg(  \frac{1}{ k_0^2-\omega_1^2}
   {-}\frac{1}{k_0^2-W_2^2}\bigg) \,, \notag
\end{align}
or 
  \begin{align} 
  D_{\mu\nu}(z)  &=\frac{\gamma}{   g^2\gamma^2   }\int \frac{\mathrm{d}^2\vec{k}}{(2\pi)^2}
    \oint_\mathcal{C}\frac{\mathrm{d}k_0}{(2\pi i)}e^{-i k\cdot z} \notag  \\ &   \times T_{\mu\nu}(k)
    \bigg( \frac{1}{ (k^2-m_1^2)(k^2-M_2^2)}    \bigg) \,.
\end{align}

Finally we arrive at 
\begin{align}
    D_{\mu\nu}(z)&=-\gamma\int \frac{\mathrm{d}^2\vec{k}}{(2\pi)^2}
    \oint_\mathcal{C}\frac{\mathrm{d}k_0}{(2\pi i )}\frac{  T_{\mu\nu}(k)}{k^2-\gamma^2\mathcal{M}^2} e^{-i k\cdot z}\,.
\end{align}
and again integrating
\begin{align}
    D_{\mu\nu}(z)
    &=-\frac{\gamma}{(m_1^2-M_2^2)}\int \frac{\mathrm{d}^2\vec{k}}{(2\pi)^2}  \bigg[ 
    \frac{T_{\mu\nu}(\omega_1,\vec{k})}  {2\omega_1}   e^{-iz_0 \omega_1}
   \notag \\
   &-\frac{  T_{\mu\nu}(-\omega_1,\vec{k})}  {2\omega_1}e^{iz_0 \omega_1}-\frac{T_{\mu\nu}(W_2,\vec{k})}{2W_2}e^{-iz_0 W_2} 
  \notag  \\ &+
  \frac{T_{\mu\nu}(-W_2,\vec{k})}{2W_2}e^{iz_0 W_2} \bigg]e^{i\vec{k}\cdot\vec{z}}\,.
\end{align}
We will use the following relations 
\begin{align}
     T_{\mu\nu}(k)\vert_{k_0=\omega_1}&=\eta_{\mu\nu}-\frac{k_\mu k_\nu}{m_1^2}\vert_{k_0=\omega_1
     }\notag \\   &   \hspace{4em} - i\epsilon_{\mu\beta\nu}\frac{k^\beta}{m_1}\vert_{k_0=\omega_1} \,, \\
     T_{\mu\nu}(k)\vert_{k_0=-\omega_1}&=\eta_{\mu\nu}-\frac{k_\mu k_\nu}{m_1^2}\vert_{k_0=-\omega_1} \notag \\   &   \hspace{4em}
     - i\epsilon_{\mu\beta\nu}\frac{k^\beta}{m_1}\vert_{k_0=-\omega_1}\,, \\
       T_{\mu\nu}(k)\vert_{k_0=W_2}&=\eta_{\mu\nu}-\frac{k_\mu k_\nu}{M_2^2}\vert_{k_0=W_2}
       \notag \\   &   \hspace{4em} +
       i\epsilon_{\mu\beta\nu}\frac{k^\beta}{M_2}\vert_{k_0=W_2} \,, \\
     T_{\mu\nu}(k)\vert_{k_0=-W_2}&=\eta_{\mu\nu}-\frac{k_\mu k_\nu}{M_2^2}\vert_{k_0=-W_2}
     \notag \\   &   \hspace{4em} +
     i\epsilon_{\mu\beta\nu}\frac{k^\beta}{M_2}\vert_{k_0=-W_2} \,,
\end{align}
where we have used~\eqref{lambda+} and~\eqref{lambda-}.

We start with the $00$ contributions 
\begin{align}
  &  \frac{T_{00}(\omega_1,\vec{k})}  {2\omega_1}  
  e^{-iz_0 \omega_1}-\frac{  T_{00}(-\omega_1,\vec{k})}  {2\omega_1}e^{iz_0 \omega_1} \notag \\
    & \hspace{4em}=-\frac{i}  {\omega_1}\sin(z_0 \omega_1)\bigg(1-\frac{\omega_1^2}{m_1^2}\bigg)\,,
\\
&-\frac{T_{00}(W_2,\vec{k})}{2W_2}e^{-iz_0 W_2} +
  \frac{T_{00}(-W_2,\vec{k})}{2W_2}e^{iz_0 W_2} \notag \\
  &  \hspace{4em} =\frac{i}  {W_2}\sin(z_0 W_2)\bigg(1-\frac{W_2^2}{M_2^2}\bigg)\,.
\end{align}
For the contributions to $I_{0i}$, we have 
\begin{align}
     &\frac{T_{0i}(k)\vert_{k_0=\omega_1}}{2\omega_1}e^{-iz_0\omega_1} -\frac{ T_{0i}(k)\vert_{k_0=-\omega_1}}{2\omega_1}e^{iz_0\omega_1} \notag \\
       &\hspace{2em}  =-\frac{ k_i}{m_1^2}\cos(z_0\omega_1)-\frac{1}{\omega_1}\epsilon_{ji}\frac{k^j}{m_1}\sin(z_0\omega_1)\,,
 \end{align}  
 and
\begin{align}
    &-\frac{T_{0i}(k)\vert_{k_0=W_2}}{2W_2}e^{-iz_0W_2} +\frac{ T_{0i}(k)\vert_{k_0=-W_2}}{2W_2}e^{iz_0W_2} \notag \\
    & \hspace{2em} =\frac{ k_i}{M_2^2}\cos(z_0W_2)-\frac{1}{W_2}\epsilon_{ji}\frac{k^j}{M_2}\sin(z_0W_2)\,,
\end{align}
where we defined $\epsilon_{0ij}=\epsilon_{ij}$.

The last ones, are the contributions from the indices $ij$ 
\begin{align}
    &\frac{T_{ij}(\omega_1,\vec{k})}  {2\omega_1}   e^{-iz_0 \omega_1}
  -\frac{  T_{ij}(-\omega_1,\vec{k})}  {2\omega_1}e^{iz_0 \omega_1}\notag \\
 & =i\frac{\delta_{ij}}  {\omega_1}  \sin(z_0 \omega_1)+i\frac{k_i k_j}{m_1^2
 \omega_1}  \sin(z_0 \omega_1) \notag \\ & \hspace{4em} + i\epsilon_{ij }\frac{1}{m_1}   \cos(z_0 \omega_1)\,,
\end{align}
and
\begin{align}
  &-\frac{T_{ij}(W_2,\vec{k})}{2W_2}e^{-iz_0 W_2} 
  \notag  +
  \frac{T_{ij}(-W_2,\vec{k})}{2W_2}e^{iz_0 W_2}  \notag \\
    &= -i\frac{ \delta_{ij}}{W_2}\sin(z_0 W_2\big) -i\frac{k_i k_j}{M_2^2W_2} \sin(z_0 W_2)
   \notag \\ & \hspace{4em} +i \epsilon_{ij} \frac{1}{M_2}\cos(z_0W_2)\,.
\end{align}
Note that the elements have the expected symmetry of $z\to -z$
and swapping $\mu \to \nu$.

Each element can be written as
\begin{subequations}
\label{D-general}
\begin{align}
    D_{00}(z)
    &=\frac{-i\gamma}{(m_1^2-M_2^2)}\notag \\
    &\times\int \frac{\mathrm{d}^2\vec{k}}{(2\pi)^2} 
    \vert\vec{k}\vert^2\bigg(\frac{\sin(z_0 \omega_1)}  {m_1^2\omega_1}
    -\frac{\sin(z_0 W_2)}  {M_2^2W_2} \bigg)e^{i\vec{k}\cdot\vec{z}} \,,
\end{align}
\begin{align}
    D_{0i}(z)
    &=\frac{\gamma}{(m_1^2-M_2^2)}\notag \\
    &\times\int \frac{\mathrm{d}^2\vec{k}}{(2\pi)^2}  \bigg[ \bigg(\frac{\cos(z_0\omega_1)}{m_1^2}
    -\frac{\cos(z_0W_2)}{M_2^2} \bigg)k_i \notag \\ &+\bigg(\frac{\sin(z_0\omega_1)}{m_1\omega_1}
    +\frac{\sin(z_0W_2)}{M_2W_2} \bigg)\epsilon_{ij}k_j\bigg]e^{i\vec{k}\cdot\vec{z}}\,,
\end{align}
\begin{align}
    D_{ij}(z)
    &=\frac{-i\gamma}{(m_1^2-M_2^2)}\notag \\
    &\times \int \frac{\mathrm{d}^2\vec{k}}{(2\pi)^2}  \bigg[   \bigg(\frac{\sin(z_0 \omega_1)}  {\omega_1} 
    -\frac{\sin(z_0 W_2 )}{W_2}\bigg)\delta_{ij}\notag \\ &+\bigg(\frac{ \sin(z_0 \omega_1)}  {m_1^2\omega_1} 
    -\frac{\sin(z_0 W_2)}{M_2^2W_2}\bigg)k_i k_j\notag \\
    &+  \bigg(\frac{1}{m_1}   \cos(z_0 \omega_1)+\frac{1}{M_2}\cos(z_0W_2)\bigg)
    \epsilon_{ij}\bigg]e^{i\vec{k}\cdot\vec{z}}\,.
\end{align}
\end{subequations}

Due to Lorentz covariance, the commutator can be evaluated in any convenient frame. 
Since we are interested in the spacelike region $z^2 = (x - y)^2 < 0$, 
we choose a frame where $z_0 = 0$. 
For completeness, the component $D_{00}(z)$ is also evaluated in a general frame 
in Appendix~\ref{App:A}.
 
In the frame wtih $z_0=0$ the elements reduce to
\begin{subequations}
\label{D-particular}
\begin{align}
    D_{00}(\vec{z})
    &=0\,,
\end{align}
\begin{align}
    D_{0i}(\vec{z})
    &=\frac{-\gamma}{m_1^2M_2^2}\int \frac{\mathrm{d}^2\vec{k}}{(2\pi)^2}
    k_ie^{i\vec{k}\cdot\vec{z}}\,,
\end{align}
\begin{align}
    D_{ij}(\vec{z})
    &=\frac{-i\gamma \epsilon_{ij}}{m_1M_2(m_1-M_2)}\int \frac{\mathrm{d}^2\vec{k}}{(2\pi)^2}  
    e^{i\vec{k}\cdot\vec{z}}\,.
\end{align}
\end{subequations}
We observe that $D_{ij}$ and $D_{0i}$ are proportional to 
$\delta(|\vec{z}|)/|\vec{z}|$ and its derivative, respectively. 
These expressions vanish outside the light cone, where $|\vec{z}| > 0$. 
Therefore, microcausality is preserved in the model.
\section{Final Remarks}\label{sectionVI}
In this work, we have studied the higher-derivative extension of 
three-dimensional QED
which incorporate gauge-invariant CS modifications. 
In particular, we have considered a QED with standard 
fermions, CS gauge invariant extensions including higher-order
and a
standard interaction term.
While earlier gauge extensions
were treated mostly at the tree level~\cite{Avila:2019xdn}, now we studied 
the one-loop perturbative impact of the 
presence of the higher-derivative CS term. 

Explicitly, we have found that the higher-order gauge field 
propagates through two different modes, one corresponding to a physical massive field and the other to a massive 
ghost field.
We have derived the polarization vectors for both Proca-like fields,
each corresponding to a subset $\{ \varepsilon^{(\pm)}\}$ described by transverse vectors in momentum space.
Also, we have quantized the 
gauge theory and verified the equal-time-commutation-relations between basis variables.
At this level, we have shown that the theory has an indefinite metric structure
which shows up in the commutators of creation and annihilation operators of the gauge field.

We have computed the two-point function of the spinor 
field which induces radiative correction 
that allow to redefine a Lagrangian in terms of two decoupled standard fermions, one being a ghost, which has served to compute the rest of raditive corrections.
With this decomposition, we  have computed the PPO where the complete dependence on the external momentum was obtained,
 and calculated the two- and three-point functions involving the spinor field.
 We have seen that in the gauge sector a redefinition of variables to cast the theory into the sum of Proca-like fields may be prohibited since these will eventually destroy gauge invariance. 
 Moreover, we explicitly checked that our theory is microcausal in the gauge sector. 
 The results for the spinor sector, can be used within the 
 studies of the three-dimensional effective field theories applied within the condensed matter context, 
 especially within studies of graphene which is a perfect three-dimensional model. We also leave for future work 
 the finite renormalization of the model.
\section*{Acknowledgments}
The work by AYuP has been supported by the
CNPq project No. 303777/2023-0. The research of CMR
was partially supported by Fondecyt Regular project No. 1241369; and wants 
to thank the kind hospitality at 
Universidad Nacional Autónoma de M\'exico (UNAM) where this work was finished. CR acknowledges 
support from the ANID fellowship No. 21211384 and Universidad de Concepción. AS was supported by DGAPA-UNAM under Grant No. PAPIIT-IN108123.
\appendix
\section{The calculation of $D_{00}$}\label{App:A}
Here we analyze a more general frame with $z_0 \neq 0$, starting from Eq.~\eqref{D-general}. 
In polar coordinates, defined by
\begin{eqnarray}
\mathrm{d}^2\vec{k}&=&\vert\vec{k}\vert\mathrm{d}
\vert\vec{k}\vert \mathrm{d}\theta \,, \\
    \vec{k}\cdot \vec{z}&=&\vert\vec{k}\vert 
    \vert\vec{z}\vert \cos\theta\,, \\
    \vec{k}&=&\big(\vert\vec{k}\vert \cos
    \theta,\vert\vec{k}\vert \sin \theta\big)\,, 
\end{eqnarray}
we obtain
\begin{subequations}
\begin{align}
    D_{00}(z)
    &=\frac{-i\gamma}{(m_1^2-M_2^2)}\int_0^{2\pi}\int_0^\infty \frac{\vert\vec{k}\vert^3\mathrm{d}\vert\vec{k}
    \vert \mathrm{d}\theta}{(2\pi)^2}  \\
    &\times \bigg(\frac{\sin(z_0 \omega_1)} 
    {m_1^2\omega_1} -\frac{\sin(z_0 W_2)} 
    {M_2^2W_2} \bigg)e^{i\vert\vec{k}\vert \vert\vec{z}\vert
    \cos\theta}\,,\notag 
\end{align}
\begin{align}
    D_{0i}(z)
    &=\frac{\gamma}{(m_1^2-M_2^2)}\int_0^{2\pi}\int_0^\infty \frac{\vert\vec{k}\vert\mathrm{d}\vert\vec{k}\vert \mathrm{d}\theta}{(2\pi)^2}  \\
    &\times \bigg[ \bigg(\frac{\cos(z_0\omega_1)}{m_1^2}-\frac{\cos(z_0W_2)}{M_2^2} \bigg)k_i\notag \\
    &+\bigg(\frac{\sin(z_0\omega_1)}{m_1\omega_1}+\frac{\sin(z_0W_2)}{M_2W_2} \bigg)\epsilon_{ij}k_j\bigg]e^{i\vert\vec{k}\vert \vert\vec{z}\vert \cos\theta}\,,\notag 
\end{align}
and
\begin{align}
    D_{ij}(z)
    &=\frac{-i\gamma}{(m_1^2-M_2^2)}\int_0^{2\pi}\int_0^\infty \frac{\vert\vec{k}\vert\mathrm{d}\vert\vec{k}\vert
    \mathrm{d}\theta}{(2\pi)^2} \\
    &\times \bigg[   \bigg(\frac{\sin(z_0 \omega_1)} 
    {\omega_1}  -\frac{\sin(z_0 W_2 )}{W_2}\bigg)\delta_{ij}\notag  \\
    &+\bigg(\frac{ \sin(z_0 \omega_1)}  {m_1^2\omega_1} 
    -\frac{\sin(z_0 W_2)}{M_2^2W_2}\bigg)k_i k_j\notag \\
    &+  \bigg(\frac{1}{m_1}   \cos(z_0 \omega_1)
    +\frac{1}{M_2}\cos(z_0W_2)\bigg)
    \epsilon_{ij}\bigg]e^{i\vert\vec{k}\vert \vert\vec{z}\vert
    \cos\theta}\,.\notag 
\end{align}
\end{subequations}
Considering the following known integrals 
\begin{align}
   \int_{0}^{2\pi }\frac{\mathrm{d}\theta}{2\pi}e^{i\vert\vec{k}\vert \vert\vec{z}\vert \cos\theta}&= J_0\big(\vert\vec{k}\vert \vert\vec{z}\vert \big)  \,, \\
   \int_0^{2\pi} \frac{\mathrm{d}\theta}{2\pi}\cos\theta e^{i \vert\vec{k}\vert \vert\vec{z}\vert 
   \cos\theta}&= iJ_1\big(\vert\vec{k}\vert \vert\vec{z}\vert \big)\,, \\
   \int_0^{2\pi}\frac{\mathrm{d}\theta}{2\pi} \cos^2\theta e^{i\vert\vec{k}\vert \vert\vec{z}\vert
   \cos\theta}&=\frac{1}{2}   \left(J_0\big(\vert\vec{k}\vert \vert\vec{z}\vert \big)-J_{2}\big(\vert\vec{k}\vert
   \vert\vec{z}\vert \big) \right)\,, \\
   \int_0^{2\pi}\frac{\mathrm{d}\theta}{2\pi} \sin^2(\theta)e^{i\vert\vec{k}\vert \vert\vec{z}\vert \cos\theta}
   &=\frac{1}{2}  \left(  J_0\big(\vert\vec{k}\vert \vert\vec{z}\vert \big)+J_{2}\big(\vert\vec{k}\vert \vert\vec{z}\vert \big)  \right )\,, 
\end{align}
and using the fact that the following integrals vanish due to parity symmetry
\begin{align}
    \int_0^{2\pi}\frac{\mathrm{d}\theta}{2\pi}\sin(\theta)e^{i\vert\vec{k}\vert \vert\vec{z}\vert\cos\theta}&=0\,, \\
\int_0^{2\pi}\frac{\mathrm{d}\theta}{2\pi}\cos(\theta)\sin(\theta)e^{i\vert\vec{k}\vert \vert\vec{z}\vert \cos(\theta)}&=0 \,,
\end{align}
we are led to the following set of integrals to evaluate.
For $D_{00}(z)$ 
\begin{align}
    D_{00}(z)
    &=\frac{-i\gamma}{(m_1^2-M_2^2)}\int_0^\infty \frac{\vert\vec{k}\vert^3\mathrm{d}\vert\vec{k}\vert }{2\pi} \label{D00-general}\\
    &\times \bigg(\frac{\sin(z_0 \omega_1)}  {m_1^2\omega_1} -\frac{\sin(z_0 W_2)}  {M_2^2W_2} \bigg)J_0\big(\vert\vec{k}\vert \vert\vec{z}\vert \big) \,.  \notag
\end{align}
For $D_{0i}(z)$ we have the components 
\begin{subequations}
\begin{align}
    D_{01}(z)
    &=\frac{i\gamma}{(m_1^2-M_2^2)}\int_0^\infty \frac{\vert\vec{k}\vert^2\mathrm{d}\vert\vec{k}\vert }{2\pi}    \\
    &\times\bigg(\frac{\cos(z_0\omega_1)}{m_1^2}-\frac{\cos(z_0W_2)}{M_2^2} \bigg) J_1\big(\vert\vec{k}\vert \vert\vec{z}\vert\big) \,, \notag
\end{align}
\begin{align}
         D_{02}(z)
    &=\frac{-i\gamma}{(m_1^2-M_2^2)}\int_0^\infty \frac{\vert\vec{k}\vert^2\mathrm{d}\vert\vec{k}\vert }{2\pi}  \\
    &\times\bigg(\frac{\sin(z_0\omega_1)}{m_1\omega_1}+\frac{\sin(z_0W_2)}{M_2W_2} \bigg) J_1\big(\vert\vec{k}\vert \vert\vec{z}\vert\big) \notag \,.
\end{align}
\end{subequations}

Finally for the $D_{ij}(z)$ components we have:
\begin{subequations}
\begin{align}
    D_{11}(z)
    &=\frac{-i\gamma}{(m_1^2-M_2^2)}\int_0^\infty \frac{\vert\vec{k}\vert\mathrm{d}\vert\vec{k}\vert }{2\pi} \notag \\
    &\times \bigg[   \bigg(\frac{\sin(z_0 \omega_1)}  {\omega_1}  -\frac{\sin(z_0 W_2 )}{W_2}\bigg)J_0\big(\vert\vec{k}\vert\vert\vec{z}\vert\big)\notag \\
    &+\frac{\vert\vec{k}\vert^2}{2}\bigg(\frac{ \sin(z_0 \omega_1)}  {m_1^2\omega_1}  -\frac{\sin(z_0 W_2)}{M_2^2W_2}\bigg) \notag \\
    &\times \big(J_0\big(\vert\vec{k}\vert \vert\vec{z}\vert \big)-J_{2}\big(\vert\vec{k}\vert \vert\vec{z}\vert \big)\big) \bigg]\,,
\end{align}
\begin{align}
        D_{12}(z)
    &=\frac{-i\gamma}{(m_1^2-M_2^2)}\int_0^\infty \frac{\vert\vec{k}\vert\mathrm{d}\vert\vec{k}\vert }{2\pi}  \notag \\
    &\times\bigg(\frac{\cos(z_0 \omega_1)}{m_1}   +\frac{\cos(z_0W_2)}{M_2}\bigg) J_0\big(\vert\vec{k}\vert \vert\vec{z}\vert \big) \,,
\end{align}
\begin{align}
         D_{21}(z)
    &=\frac{i\gamma}{(m_1^2-M_2^2)}\int_0^{2\pi}\int_0^\infty \frac{\vert\vec{k}\vert\mathrm{d}\vert\vec{k}\vert }{2\pi}  \notag \\
    &\times \bigg(\frac{\cos(z_0 \omega_1)}{m_1}   +\frac{\cos(z_0W_2)}{M_2}\bigg) J_0\big(\vert\vec{k}\vert \vert\vec{z}\vert \big)\,,
\end{align}
\begin{align}
     D_{22}(z)
    &=\frac{-i\gamma}{(m_1^2-M_2^2)}\int_0^\infty \frac{\vert\vec{k}\vert\mathrm{d}\vert\vec{k}\vert }{2\pi}  \notag \\
    &\times \bigg[   \bigg(\frac{\sin(z_0 \omega_1)}  {\omega_1}  -\frac{\sin(z_0 W_2 )}{W_2}\bigg)J_0\big(\vert\vec{k}\vert \vert\vec{z}\vert\big)\notag \\
    &+\frac{\vert\vec{k}\vert^2}{2}\bigg(\frac{ \sin(z_0 \omega_1)}  {m_1^2\omega_1}  -\frac{\sin(z_0 W_2)}{M_2^2W_2}\bigg)\notag \\
    &\times \big(J_0\big(\vert\vec{k}\vert \vert\vec{z}\vert\big)+J_2\big(\vert\vec{k}\vert \vert\vec{z}\vert\big)\big)\bigg]\,.
\end{align}
\end{subequations}
To illustrate how the case with $z_0 \neq 0$ leads to the same conclusion as the $z_0 = 0$ frame, 
we explicitly compute the $D_{00}(z)$ integral given in Eq.~\eqref{D00-general}. 
In this case, we encounter an integral of the form:
\begin{eqnarray}
    I(z)=\int_0^\infty \frac{k^2\sin\big(z_0\omega(k)\big)}
    {\omega(k)}J_0(k z)k\mathrm{d}k\,,
\end{eqnarray}
for $\omega=\omega_1,W_2$. We know that the Bessel functions satisfy the ODE
\begin{align}
    x^2 \frac{\mathrm{d}^2J_n(x)}{\mathrm{d}x^2}+x\frac{\mathrm{d}J_n(x)}{\mathrm{d}x}+(x^2-n^2)J_n(x)=0\,,
\end{align}
that in our particular case means 
\begin{eqnarray}
    \bigg(\frac{\mathrm{d}^2}{\mathrm{d}z^2}+\frac{1}{z}\frac{\mathrm{d}}{\mathrm{d}z}\bigg)J_0(kz)=-k^2J_0(kz) \,.
\end{eqnarray}
By using this expression we can write the integral as
\begin{align}
    I(z)=-\Big( \partial_z^2+\frac{1}{z}\partial_z \Big)\bigg[ \int_0^\infty
    \frac{\sin\big(z_0\omega(k)\big)}{\omega(k)}J_0(kz)k\mathrm{d}k\bigg] \,.
\end{align}
Now by changing the integration variable as $\tau=\omega(k),\mathrm{d}\tau=\frac{k\mathrm{d}k}{\omega(k)}$ then we obtain
\begin{align}
    &\int_0^\infty \frac{\sin\big(z_0\omega(k)\big)}{\omega(k)}J_0(kz)k\mathrm{d}k\notag \\
    &= \int_m^\infty \sin\big(z_0\tau\big)J_0(z\sqrt{\tau^2-m^2})\mathrm{d}\tau \,.
\end{align}

From \cite{Gradshteyn:1943cpj} (item 6.677-1), 
the previous integral has the form
\begin{widetext}
\begin{align}
    \int_a^\infty J_0(b\sqrt{x^2-a^2})\sin(cx)\mathrm{d}x=\left\lbrace
    \begin{matrix} 0 &,0<c<b\\ \frac{\cos\big(a\sqrt{c^2-b^2}\big)}{\sqrt{c^2-b^2}}&,0<b<c \end{matrix}\right.
\end{align}
In our case $a=m$, $b=z$ and $c=z_0$, thus 
\begin{align}
    \int_m^\infty \sin\big(z_0\tau\big)J_0(z\sqrt{\tau^2-m^2})
    \mathrm{d}\tau=\left\lbrace \begin{matrix} 0 &,0<z_0<z\\
    \frac{\cos\big(m\sqrt{z_0^2-z^2}\big)}{\sqrt{z_0^2-z^2}}&,0<z<z_0 \end{matrix}\right. 
\end{align}
\end{widetext}
We conclude that outside the lightcone the integral $I(z)$ vanishes. The other integrals can 
be calculated in the same way.

\end{document}